\documentclass[journal,compsoc]{IEEEtran}

\ifCLASSINFOpdf
\else
\fi

\usepackage{times}
\usepackage{subfigure}
\usepackage{amsmath}
\usepackage{amsfonts}
\usepackage{url}
\usepackage{graphicx}
\usepackage{parskip}
\usepackage{cite}
\usepackage{amssymb}
\usepackage{color, soul}

\ifCLASSOPTIONcompsoc

\else
\fi

\ifCLASSINFOpdf

\else

\fi

\begin{document}

\title{Image Retargeting by Content-Aware Synthesis}

\author{Weiming Dong~\IEEEmembership{Member,~IEEE,} Fuzhang~Wu, Yan~Kong, Xing~Mei~\IEEEmembership{Member,~IEEE,}
        Tong-Yee~Lee~\IEEEmembership{Senior Member,~IEEE,}
        and~Xiaopeng~Zhang,~\IEEEmembership{Member,~IEEE}
\IEEEcompsocitemizethanks{\IEEEcompsocthanksitem Weiming Dong, Fuzhang Wu, Yan Kong, Xing Mei and Xiaopeng Zhang are with NLPR-LIAMA, Institute of Automation, Chinese Academy of Sciences, Beijing, China. E-mail: \{Weiming.Dong, Xiaopeng.Zhang\}@ia.ac.cn.
\IEEEcompsocthanksitem Tong-Yee Lee is with National Cheng-Kung University, Taiwan. E-mail: tonylee@mail.ncku.edu.tw.}
\thanks{}}

\markboth{IEEE Transactions on Visualization and Computer Graphics,~Vol.~XX, No.~XX, June~2014}%
{Dong \MakeLowercase{\textit{et al.}}: Image Retargeting by Content-Aware Synthesis}

\IEEEcompsoctitleabstractindextext{
\begin{abstract}

Real-world images usually contain vivid contents and rich textural details, which will complicate the manipulation on them. In this paper, we design a new framework based on content-aware synthesis to enhance content-aware image retargeting. By detecting the textural regions in an image, the textural image content can be \textit{synthesized} rather than simply distorted or cropped. This method enables the manipulation of textural \& non-textural regions with different strategy since they have different natures. We propose to retarget the textural regions by content-aware synthesis and non-textural regions by fast multi-operators. To achieve practical retargeting applications for general images, we develop an automatic and fast texture detection method that can detect multiple disjoint textural regions. We adjust the saliency of the image according to the features of the textural regions. To validate the proposed method, comparisons with state-of-the-art image targeting techniques and a user study were conducted. Convincing visual results are shown to demonstrate the effectiveness of the proposed method.

\end{abstract}

}

\maketitle

\IEEEdisplaynotcompsoctitleabstractindextext
\IEEEpeerreviewmaketitle

\section{Introduction}

\IEEEPARstart{I}{mage} retargeting has retained in the front rank of most widely-used digital media processing techniques for a long time. To adapt raw image material for a specific use, there are often the needs of achieving a target resolution by reducing or inserting image content. To protect certain important areas, some methods~\cite{Avidan:07,Wang:08,Panozzo:2012,Lin:2013:PBI} used significance maps based on local low-level features such as gradient, dominant colors, and entropy. However, high-level semantics also play an important role in human's image perception, so usually it is necessary to better understand the content of an image to help to choose a more feasible scheme for retargeting operation. Moreover, as found in \cite{Rubinstein:2010}, viewers are more sensitive to deformation than to image area loss. Therefore for some examples it is better to summarize the content rather than distort/warp or crop the origin images~\cite{Simakov:08,Wu:2010,Dong:2014:SBI}.

\begin{figure*}[t!]
\centering
 \includegraphics[width=0.98\linewidth]{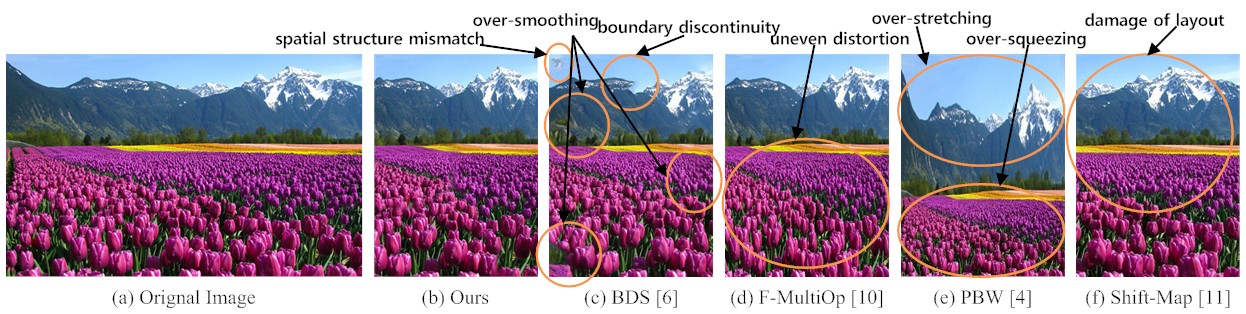}
 \vskip -2mm
 \caption{\protect \small By spatially synthesizing textural regions in (a) the input image, our method (b) can retain the remained objects without over-smoothing ((c)), spatial structure mismatch ((c)), boundary discontinuity ((c)), uneven distortion ((d)), over-squeezing/over-stretching ((e)), or damage of layout ((f)). Input $555\times 347$, output $256 \times 347$. $65.45\%$ users favour our result.}
 \vskip -3mm
  \label{fig:tulip04}
\end{figure*}

\begin{figure*}[t!]
\centering
\subfigure[Original Image]{
  \includegraphics[width=0.26\linewidth]{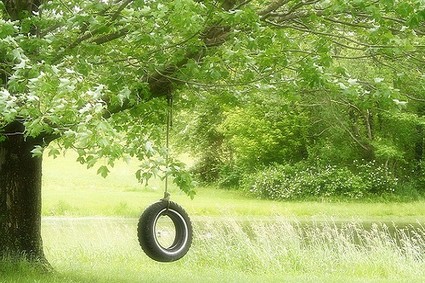}
  \label{fig:tire_o}
  }
  \hskip -1.8mm
  \subfigure[Ours]{
  \includegraphics[width=0.1352\linewidth]{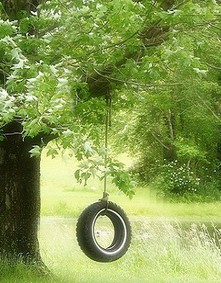}
  \label{fig:tire_texd}
  }
  \hskip -1.8mm
  \subfigure[AAD~\cite{Panozzo:2012}]{
  \includegraphics[width=0.1352\linewidth]{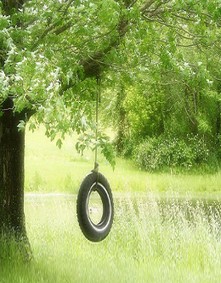}
  \label{fig:tire_aad}
  }
  \hskip -1.8mm
  \subfigure[BDS~\cite{Simakov:08}]{
  \includegraphics[width=0.1352\linewidth]{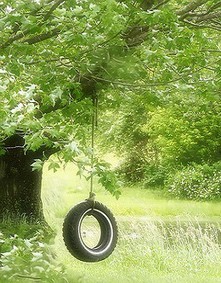}
  \label{fig:tire_bd}
  }
  \hskip -1.8mm
  \subfigure[F-MultiOp~\cite{Dong:2012}]{
  \includegraphics[width=0.1352\linewidth]{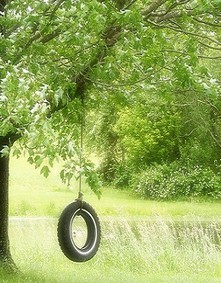}
  \label{fig:tire_mod}
  }
  \hskip -1.8mm
  \subfigure[Shift-Map~\cite{Pritch:09}]{
  \includegraphics[width=0.1352\linewidth]{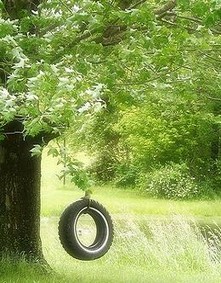}
  \label{fig:tire_sm}
  }
  \vskip -2mm
 \caption{\protect \small Most contents of the original image are textures. Input $500\times 333$, output $260 \times 333$. $63.64\%$ users favour our result.}
 \vskip -2mm
  \label{fig:tire}
\end{figure*}

\begin{figure*}[t!]
\centering
\subfigure[Original Image]{
  \includegraphics[width=0.268\linewidth]{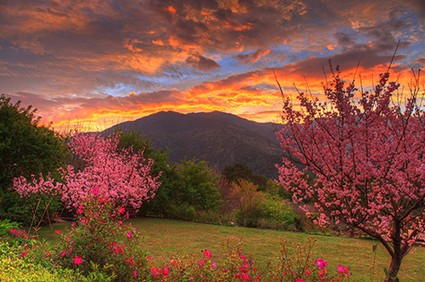}
  \label{fig:field23_o}
  }
  \hskip -1.8mm
  \subfigure[Ours]{
  \includegraphics[width=0.134\linewidth]{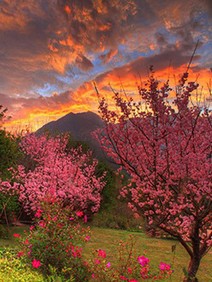}
  \label{fig:field23_texd}
  }
  \hskip -1.8mm
  \subfigure[BDS~\cite{Simakov:08}]{
  \includegraphics[width=0.134\linewidth]{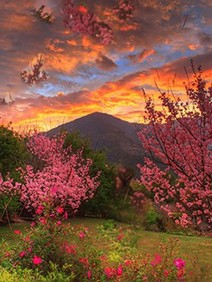}
  \label{fig:field23_bd}
  }
  \hskip -1.8mm
  \subfigure[F-MultiOp~\cite{Dong:2012}]{
  \includegraphics[width=0.134\linewidth]{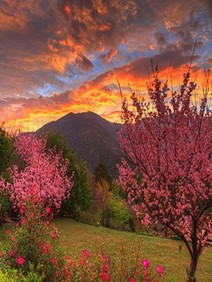}
  \label{fig:field23_mod}
  }
  \hskip -1.8mm
  \subfigure[PBW~\cite{Lin:2013:PBI}]{
  \includegraphics[width=0.134\linewidth]{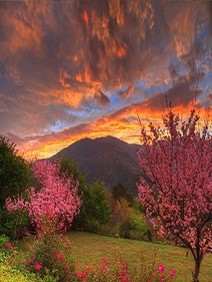}
  \label{fig:field23_pbw}
  }
  \hskip -1.8mm
  \subfigure[Shift-Map~\cite{Pritch:09}]{
  \includegraphics[width=0.134\linewidth]{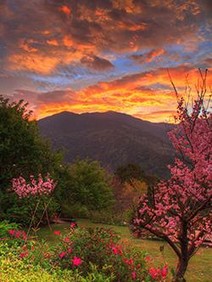}
  \label{fig:field23_sm}
  }
    \vskip -2mm
 \caption{\protect \small Most contents of the original image are textures. Input $500\times 332$, output $250 \times 332$. $78.18\%$ users favour our result.}
 \vskip -4mm
  \label{fig:field23}
\end{figure*}

Although many retargeting methods have been proposed, a few noticeable and critically influencing issues still endure. They are mostly related to complexity of textural patterns in many natural images. Previous retargeting techniques attempt to modify the image without noticing the properties of textural regions, and may easily result in apparent visual artifacts, such as over-smoothing (Figs.~\ref{fig:tulip04}(c), \ref{fig:tire_bd}, \ref{fig:field23_bd}), local boundary discontinuity (Figs.~\ref{fig:tulip04}(c), \ref{fig:tire_sm}), content spatial structure mismatch (Figs.~\ref{fig:tulip04}(c), \ref{fig:tire_bd}, \ref{fig:field23_bd}, \ref{fig:field23_sm}), uneven distortion (Figs.~\ref{fig:tulip04}(d), \ref{fig:tire_aad}, \ref{fig:tire_mod}, \ref{fig:field23_mod}), over-squeezing/over-stretching (Figs.~\ref{fig:tulip04}(e), \ref{fig:tire_aad}. \ref{fig:field23_pbw}), and damage of scene layout (Figs.~\ref{fig:tulip04}(f), \ref{fig:field23_sm}). The examples in Figs.~\ref{fig:tulip04}-\ref{fig:field23} are not special and exhibit one common problem - that is, \textit{when images contain large textural patterns, retargeting quality could be generally affected by their complexity.} Since regularity is an important high-level feature for human texture perception~\cite{Rao:1993:CGM} and texture exists in many natural images, this problem cannot be ignored.

We propose a novel content-aware synthesis algorithm to address the challenge of handling textural patterns in image retargeting. In preprocessing, the textural regions (T-regions) of the input image are automatically detected based on local variation measures and each pixel in a T-region is assigned a significance value. In the retargeting process, the input image is first retargeted to the target size by fast multi-operators (F-MultiOp). Then, the T-regions are regenerated by \textit{synthesis}, which arranges sample patches with respect to the neighborhood metric and patch position information (Figs.~\ref{fig:tulip04}(b), \ref{fig:tire_texd}, \ref{fig:field23_texd}). The patches with higher significance values have higher probability to appear in the result. With the content-based information and texture synthesis technique, the proposed approach can better protect both the local shape of the texture elements (texels) and the global visual appearance of the T-regions than previous image retargeting methods.

Compared with recent studies on image retargeting, the major contributions of the proposed approach are as follows:
\begin{itemize}
\item A fast and automatic method to detect the T-regions in an image. This process makes it possible for retargeting operation to treat the T-regions and NT-regions (non-textural regions) with different strategies.
\item A novel texture saliency detection method to generate significance map in a T-region, which is based on both color and texture features.
\item A synthesis-enhanced image retargeting approach is proposed to ease unpleasant visual distortions caused by seam carving, warping or scaling to overall texels in T-regions. Thus, our approach can yield better results in terms of texture element (texel) shape and preservation of globally varying effect compared with related approaches.
\end{itemize}

To compare with the state-of-the-arts image retargeting methods, we construct a new benchmark image set and conduct a user study to demonstrate the effectiveness of our framework.

\vspace{-4mm}
\section{Related Works}
\label{sec:related}

\vspace{-2mm}
\subsection{Image Retargeting}

\vspace{-3mm}
Numerous content-aware image retargeting techniques have recently been proposed. Cropping has been widely used to eliminate the unimportant information from the image periphery or improve the overall composition~\cite{Zhang:2013:PGT,Yan:2013:LCA,Zhang:2014:WSP}. Seam carving methods iteratively remove a seam in the input image to preserve visually salient content~\cite{Avidan:07,Rubinstein:08}. A seam is a continuous path with minimum significance. Multi-operator algorithms combine seam carving, homogeneous scaling and cropping to optimally resize images~\cite{Rubinstein:09,Dong:2009c,Dong:2012}. Pritch et al.~\cite{Pritch:09} introduced Shift-Map that removed or added band regions instead of scaling or stretching images. For many cases these \textit{discrete} approaches can generate pleasing results, however, the seam removal may cause discontinuous artifacts, and cropping is unsuitable for the case when there are visually salient contents near the borders of images.

\vspace{-1mm}
\textit{Continuous} retargeting methods have been realized through image warping or mapping by using several deformation and smoothness constraints~\cite{Gal:06,Wolf:07,Wang:08,Zhang:08,Krahenbuhl:2009,Guo:2009,Liang:2013:OSF}. A finite element method has also been used to formulate image warping~\cite{Kaufmann:2013:FEI}. Recent continuous retargeting methods focus on preserving local structures. Panozzo et al.~\cite{Panozzo:2012} minimize warping energy in the space of axis-aligned deformations to avoid harmful distortions. Chang et al.~\cite{Chang:2012:LAI} couple mesh deformations with similarity transforms for line features to preserve line structure properties. Lin et al.~\cite{Lin:2013:PBI} present a patch-based scheme with an extended significance measurement to preserve shapes of both visual salient objects and structural lines. These approaches perform well on shape preservation of salient objects but often over-squeeze or over-stretch the T-regions to distort all texels since T-regions are usually not salient in the whole image.
\begin{figure*}[t!]
\centering
\subfigure[Some examples used in this paper]{
  \includegraphics[width=0.47\linewidth]{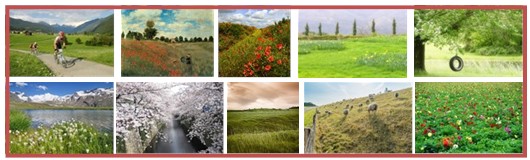}
  \label{fig:exp}
  }
  \hskip -2.5mm
  \subfigure[Some examples in RetargetMe benchmark~\cite{Rubinstein:2010}]{
  \includegraphics[width=0.504\linewidth]{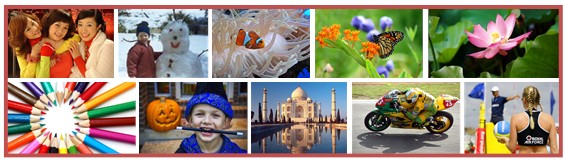}
  \label{fig:exp_bench}
  }
  \vskip -3mm
  \caption{\protect \small Different with most of the images in the RetargetMe benchmark, our exemplar images all contain large textural regions.}
  \vskip -4mm
  \label{fig:examples}
\end{figure*}
\begin{figure*}[t!]
\centering
\includegraphics[width=0.95\linewidth]{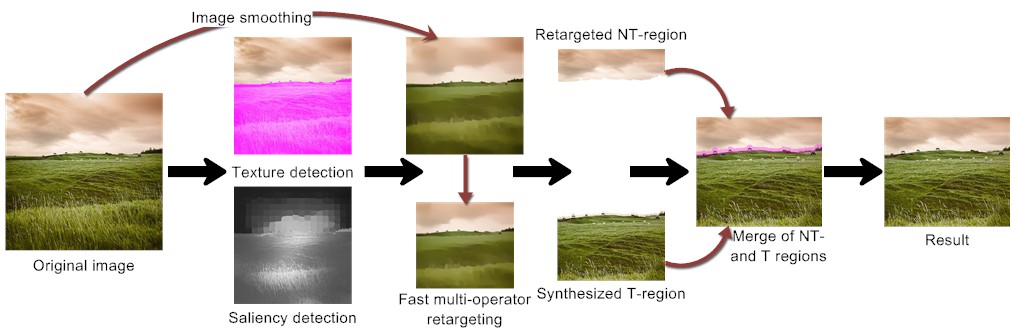}
\vskip -2mm
\caption{\protect \small Framework of our method. Input resolution $400\times 400$. Target resolution $240 \times 210$.}
\vskip -4mm
\label{fig:workflow}
\end{figure*}

\vspace{-3mm}
\textit{Summarization}-based retargeting approaches eliminate repetitive patches instead of individual pixels and preserve patch coherence between the source and target image during retargeting~\cite{Simakov:08,Cho:08,Barnes:09}. These techniques measure patch similarity and select patch arrangements that fit together well to change the size of an image. However, due to the lack of enough content information, the major drawback of such methods is that the globally visual effect may be discarded and some regions may be over-smoothed when the target size is small.

High level image content informations are analyzed and integrated in some recent summarization approaches. For example, Wu et al.~\cite{Wu:2010} detect the corresponding lattice of a symmetry image region and retarget it by trimming the lattice. Basha et al.~\cite{Basha:2013:SSC} employ depth information to maintain geometric consistence when retargeting stereo images by seam carving. Lin et al.~\cite{Lin:2014:OCW} utilize the object correspondences in the left and right images of a stereoscopic image in retargeting, which allows the generation of an object-based significance map and the consistent preservation of objects during warping. Dong et al.~\cite{Dong:2014:SBI} detect similar objects in the input image and then use object carving to achieve a natural retargeting effects with minimum object saliency damage. There also exist a few efforts to deal with textures for better retargeting. Kim and Kim~\cite{Kim:2011:TAS} exploit the higher order statistics of the diffusion space to define a reliable image importance map, which can better preserve the salient object when it is located in front of a textural background. This approach does not consider how to preserve the visual effects of textural regions. Zhang and Kuo~\cite{Zhang:2012:RAT} resize the salient and irregular regions by warping and re-synthesize the regular regions. However, in \cite{Zhang:2012:RAT} the authors did not address what situations the regularity detection algorithm works which questions its robustness. On the other hand, the synthesis algorithm they used can only deal with isotropic textures which is not fit for most natural images with vivid anisotropic texture regions.

\vspace{-6mm}
\subsection{Texture Detection and Synthesis}

\vspace{-3mm}
The adaptive integration of the color and texture attributes in image segmentation is one of the most investigated topics of research in computer vision (surveyed in \cite{Ilea:2011}). However, most of the image segmentation algorithm do not clearly illustrate the type of each region (textural or non-textural) in the result. Targhi et al.~\cite{Targhi:2006} present a fast texture descriptor based on LU transform, but how to determine if a pixel is texture or non-texture according to the feature values is not discussed. Bergman et al.~\cite{Bergman:2007} present an intuitive texture detection method which is based on contrast and disorganization measurements of image blocks. The method is not effective on noisy images which tend to have decreasing contrast and often generate many disjoint areas. Todorovic and Ahuja~\cite{Todorovic:2009} formulate the detection of texture subimages as identifying modes of the pdf of region descriptors. However, the method is not efficient (5 minutes for a $512 \times 512$ image) for practical image retargeting applications.

\vspace{-1mm}
Texture synthesis is a general example-based methodology for synthesizing similar phenomena~\cite{Wei:2009}. However, the basic MRF-based scheme in most existing texture synthesis methods cannot adequately handle the globally visual variation of texels, such as perspective appearance and semantic content distribution. Dong et al.~\cite{Dong:2008} present a perspective-aware texture synthesis algorithm by analyzing the size variation of the texel, but verbatim copying artifacts also often appear in their results. On the other hand, common texture synthesis algorithms are designed for enlargement and can not be directly used for image retargeting applications. Wei~\cite{Wei:2008} presents inverse texture synthesis approach to generate a smaller example from a large input texture. However, for globally varying textures, the output quality of this approach usually depends on the accuracy of the original map. Therefore, if applied to normal T-region retargeting, it will easily lose the globally visual variation or damage the local content continuity of the original image in the result.

\vspace{-4mm}
\section{Retargeting by Synthesis}
\label{sec:retarget}

\vspace{-2mm}
\subsection{System Overview}

\vspace{-2mm}
Some standard examples studied in our work are shown in Fig.~\ref{fig:exp}. Different from most examples in the RetargetMe benchmark, our images all contain one or more large textural regions, which bring new challenges to image retargeting. Previous methods can well preserve the shape of one or more salient objects in the retargeting results but often omit the "background" textures which also play important roles in most natural images. The shape of texels and some globally visual effects of T-regions will be easily damaged in the results. Our method will address those problems. Fig.~\ref{fig:workflow} illustrates the framework of the proposed method. In the preprocessing step, the input image is segmented into one or more T-regions and one NT-region (we treat disjoint NT-regions also as one region) by texture detection (Sec.~\ref{sec:texdet}). A hierarchical saliency detection for texture is then performed to generate a significance map for each T-region (Sec.~\ref{sec:texsal}). The significance map of the whole image is also adaptively adjusted according to the percentage of areas of T-regions. Afterwards, the input image is filtered by structure-preserving image smoothing. In the image retargeting step, fast multi-operator (F-MultiOp)~\cite{Dong:2012} method is firstly used to resize the filtered input image to the target size (Sec.~\ref{sec:initret}). The process of retargeting the smoothed image is used to guide the resizing process of the original image in order to eliminate the effect of textural details (Sec.~\ref{sec:initret}). We then re-generate the T-regions of the resulting image via the proposed content-aware synthesis operator, in order to maintain the perspective variation, content diversity, as well as the texel shapes (Sec.~\ref{sec:syn}). Finally, we refine the boundaries between T- and NT-regions by re-synthesizing the pixels of the boundary areas (Sec.~\ref{sec:merge}).
\begin{figure*}[t!]
\centering
\subfigure[Original]{
  \includegraphics[width=0.148\linewidth]{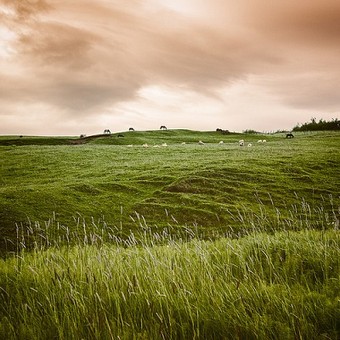}
  \label{fig:field13_tex_o}
  }
  \subfigure[RTV]{
  \includegraphics[width=0.148\linewidth]{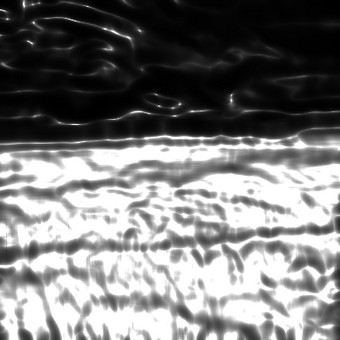}
  \label{fig:field13_tex_rtv}
  }
  \subfigure[Noisy Mask]{
  \includegraphics[width=0.148\linewidth]{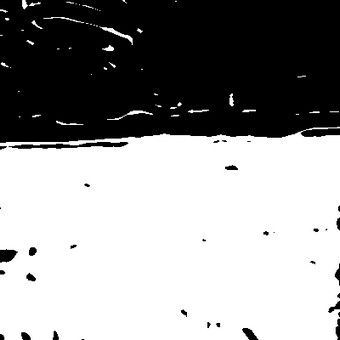}
  \label{fig:field13_tex_noisy}
  }
  \subfigure[SLIC]{
  \includegraphics[width=0.148\linewidth]{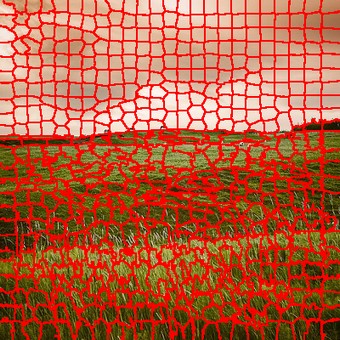}
  \label{fig:field13_tex_slic}
  }
  \subfigure[Smooth Mask]{
  \includegraphics[width=0.148\linewidth]{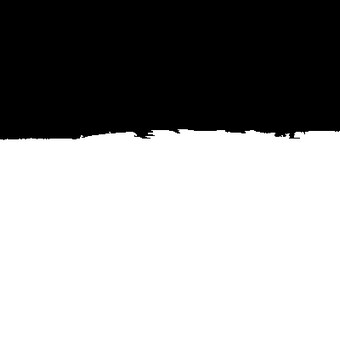}
  \label{fig:field13_tex_mask}
  }
  \subfigure[Final T-region]{
  \includegraphics[width=0.148\linewidth]{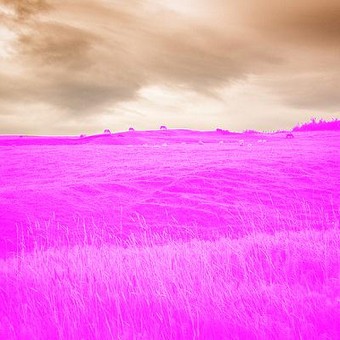}
  \label{fig:field13_tex_final}
  }
  \vskip -3mm
 \caption{\protect \small Texture detection. (e) Mask refined by utilizing super-pixels. (f) The final T-region boundary is refined by using graphcut.}
 \vskip -4mm
  \label{fig:field13_tex}
\end{figure*}

\vspace{-6mm}
\subsection{Automatic Texture Detection}
\label{sec:texdet}

\vspace{-2mm}
The first step for our image retargeting system is to locate the T-regions. Recently, local variation measures were used to smooth texture and extract structures from an image~\cite{Buades:2010:FCT,Xu:2012:SET,Karacan:2013:SIS}. However, this kind of approaches all can not provide the positional information of textures, especially for most natural images which contain both T- and NT- regions. We develop a fast texture detection method based on the measure of relative total variation (RTV). Given an input image, we first calculate the \textit{windowed total variations} $\mathcal{D}_x(p)$ (in the $x$ direction) and $\mathcal{D}_y(p)$ (in the $y$ direction) for pixel $p$, as well as the \textit{windowed inherent variations} $\mathcal{L}_x(p)$ and $\mathcal{L}_y(p)$. Details of calculating the windowed variations are described in \cite{Xu:2012:SET}. We then calculate the reliability of pixel $p$ being a texture pixel as:
\begin{equation}
R(p) = \frac{\mathcal{D}_x(p)}{\mathcal{L}_x(p) + \epsilon} + \frac{\mathcal{D}_y(p)}{\mathcal{L}_y(p) + \epsilon},
\label{equ:tex_r}
\end{equation}
where the division is an element wise operation. $\epsilon = 10^{-5}$ is used to avoid division by zero.

After calculating the reliability of each pixel, we use an iterative algorithm to set a threshold $R_T$ to determine the textural pixels. We first calculate the average reliability $R_A$ of all the pixels and use $R_T = R_A$ to separate the pixels into two parts. The pixels which $R(p) \geqslant R_T$ are set as textural pixels (T-pixels) and $R(p) < R_T$ as non-textural pixels (NT-pixels). We then calculate the average reliability of T-pixels as $R_A^T$ and the one of NT-pixels as $R_A^{NT}$. After that, we set the new threshold as $R_T' = \alpha \cdot R_A^T + (1.0 - \alpha) \cdot R_A^{NT}$, where $\alpha  = 0.5$ in all our experiments. We update $R_T = R_T'$ and repeat the above steps until $|R_T' - R_T| < \epsilon$.

We can get a noisy texture mask (see Fig.~\ref{fig:field13_tex_noisy}) after segmenting the original image into T-pixels and NT-pixels. To improve the quality of the mask, we over-segment the input image into super-pixels by SLIC~\cite{Achanta:2012:SSC} (see Fig.~\ref{fig:field13_tex_slic}). A super-pixel is labeled as texture if more than half of its pixels are labeled as texture. The smooth texture mask (see Fig.~\ref{fig:field13_tex_mask}) is further improved by graphcut in order to get more accurate boundaries for T-regions. As shown in Fig.~\ref{fig:field13_tex_final}, our algorithm can accurately detect the grassland as a T-region. Please see more texture detection results and the analysis of the accuracy of the algorithm in the supplemental material.
 \vskip -3mm
\begin{figure}[h]
\centering
\subfigure[Original]{
  \includegraphics[width=0.23\linewidth]{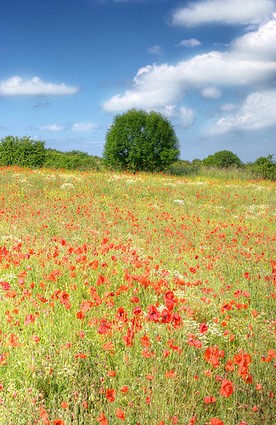}
  \label{fig:field01_o}
  }
  \hskip -2mm
  \subfigure[Our Retargeting]{
  \includegraphics[width=0.23\linewidth]{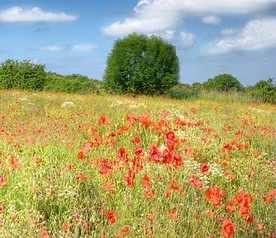}
  \label{fig:field01_texd}
  }
  \hskip -2mm
  \subfigure[Our Saliency]{
  \includegraphics[width=0.23\linewidth]{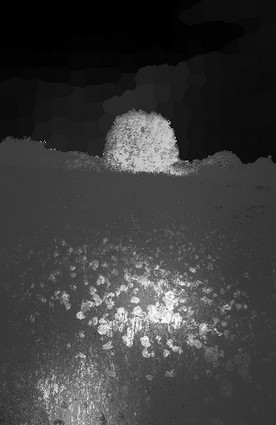}
  \label{fig:field01_sal}
  }
  \hskip -2mm
  \subfigure[\protect\cite{Cheng:2011:GCB}]{
  \includegraphics[width=0.23\linewidth]{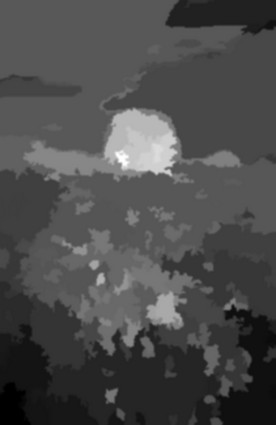}
  \label{fig:field01_sal_rc}
  }
  
  \vskip -1mm
  \subfigure[\protect\cite{Goferman:2012:CSD}]{
  \includegraphics[width=0.23\linewidth]{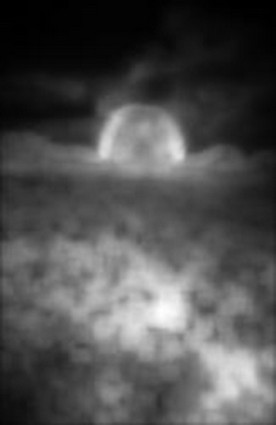}
  \label{fig:field01_tex_noisy}
  }
  \hskip -2mm
  \subfigure[\protect\cite{Yang:2013:SDG}]{
  \includegraphics[width=0.23\linewidth]{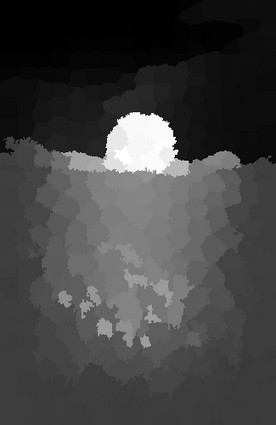}
  \label{fig:field01_sal_sdg}
  }
  \hskip -2mm
  \subfigure[\protect\cite{Yan:2013:HSD}]{
  \includegraphics[width=0.23\linewidth]{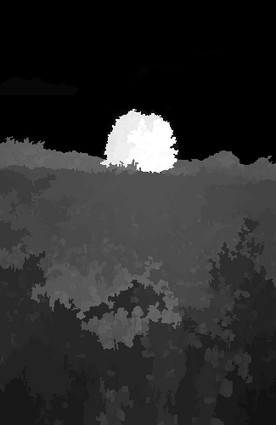}
  \label{fig:field01_sal_hsd}
  }
  \hskip -2mm
  \subfigure[\protect\cite{Margolin:2013:WMP}]{
  \includegraphics[width=0.23\linewidth]{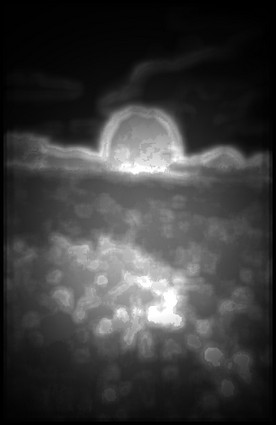}
  \label{fig:field01_sal_wmp}
  }
  
  \vskip -1mm
  \subfigure[Ours by ANN]{
  \includegraphics[width=0.23\linewidth]{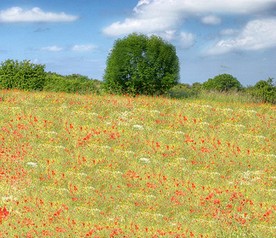}
  \label{fig:field01_texd_k}
  }
  \hskip -2mm
  \subfigure[BDS]{
  \includegraphics[width=0.23\linewidth]{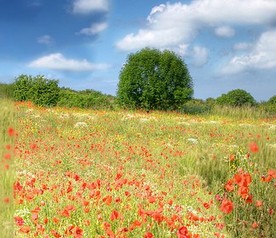}
  \label{fig:field01_bds}
  }
  \hskip -2mm
  \subfigure[PBW]{
  \includegraphics[width=0.23\linewidth]{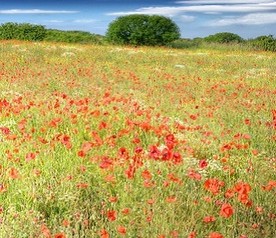}
  \label{fig:field01_pbw}
  }
  \hskip -2mm
  \subfigure[Shift-Map]{
  \includegraphics[width=0.23\linewidth]{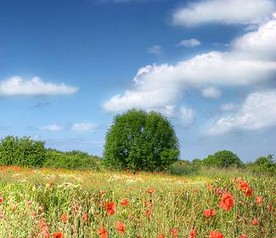}
  \label{fig:field01_sm}
  }
  \vskip -2mm
 \caption{\protect \small Qualitative comparison of saliency detection on a normal image. Previous methods fail to detect the salient dark green grass on the left-bottom. $61.82\%$ users favour our result.}
  \label{fig:field01}
  \vskip -3mm
\end{figure}
\begin{figure*}[t!]
\subfigure[Original]{
  \includegraphics[width=0.158\linewidth]{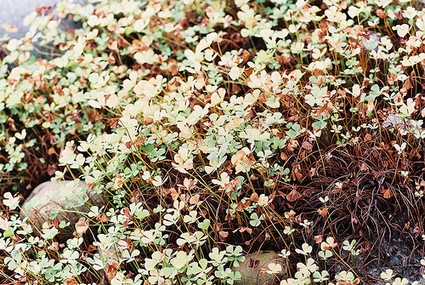}
  \label{fig:clover01_o}
  }
  \hskip 4.9mm
  \subfigure[\scriptsize Retargeting]{
  \includegraphics[width=0.079\linewidth]{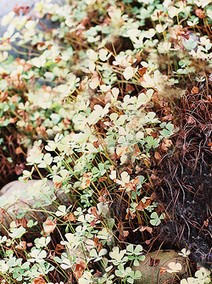}
  \label{fig:clover01_texd}
  }  
  \hskip 4.9mm
  \subfigure[Color Distinctness]{
  \includegraphics[width=0.158\linewidth]{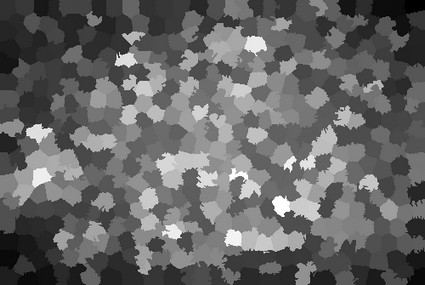}
  \label{fig:clover01_sal_col2}
  }
  \hskip -2mm
  \subfigure[Texture Distinctness]{
  \includegraphics[width=0.158\linewidth]{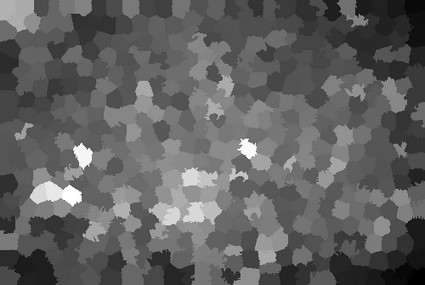}
  \label{fig:clover01_sal_tex2}
  }
  \hskip -2mm
  \subfigure[Color + Texture]{
  \includegraphics[width=0.158\linewidth]{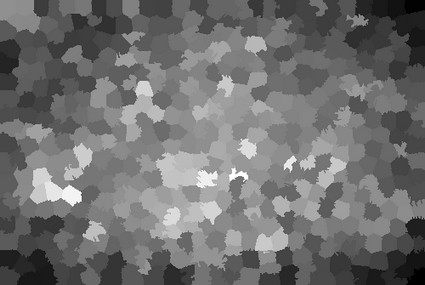}
  \label{fig:clover01_sal_texcol2}
  }
  \hskip -2mm
  \subfigure[Averaged Cue Map]{
  \includegraphics[width=0.158\linewidth]{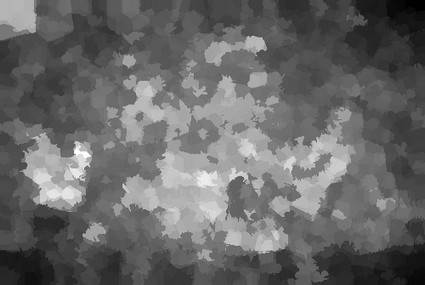}
  \label{fig:clover01_sal_texcol}
  }
  
  \vskip -1.5mm
  \subfigure[Our final saliency]{
  \includegraphics[width=0.158\linewidth]{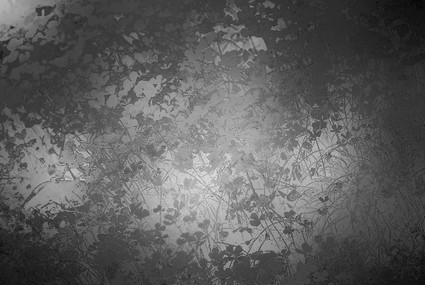}
  \label{fig:clover01_sal}
  }
  \hskip -2mm
  \subfigure[\scriptsize RC~\protect\cite{Cheng:2011:GCB}]{
  \includegraphics[width=0.158\linewidth]{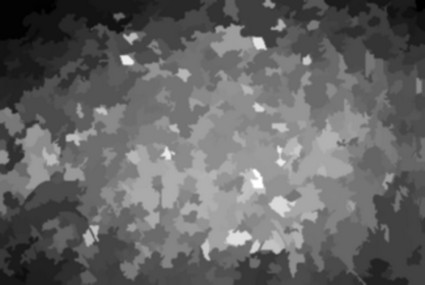}
  \label{fig:clover01_sal_rc}
  }
  \hskip -2mm
  \subfigure[\tiny CAS~\protect\cite{Goferman:2012:CSD}]{
  \includegraphics[width=0.158\linewidth]{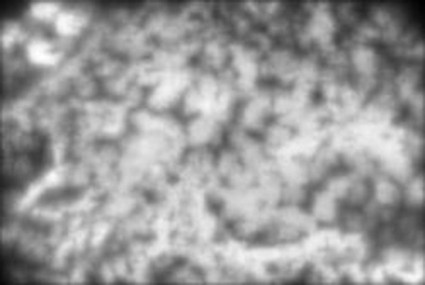}
  \label{fig:clover01_tex_noisy}
  }
  \hskip -2mm
  \subfigure[\scriptsize SDG~\protect\cite{Yang:2013:SDG}]{
  \includegraphics[width=0.158\linewidth]{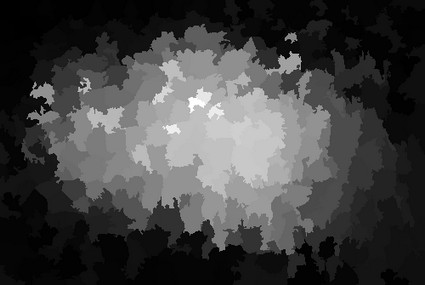}
  \label{fig:clover01_sal_sdg}
  }
  \hskip -2mm
  \subfigure[\scriptsize HSD~\protect\cite{Yan:2013:HSD}]{
  \includegraphics[width=0.158\linewidth]{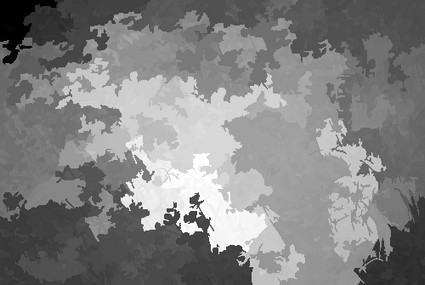}
  \label{fig:clover01_sal_hsd}
  }
  \hskip -2mm
  \subfigure[\tiny WMP~\protect\cite{Margolin:2013:WMP}]{
  \includegraphics[width=0.158\linewidth]{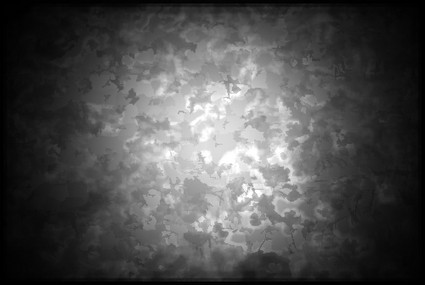}
  \label{fig:clover01_sal_wmp}
  }
  \vskip -3mm
 \caption{\protect \small Qualitative comparison of saliency detection on a pure texture image. (c), (d), (e) are calculated at Layer 2 with $M = 500$.}
  \label{fig:clover01}
  \vskip -3mm
\end{figure*}
\begin{figure*}[t!]
\centering
\subfigure[Original Image]{
  \includegraphics[width=0.21\linewidth]{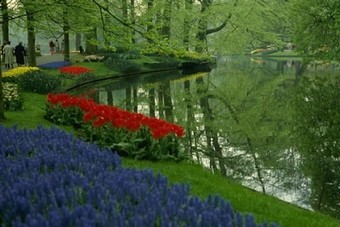}
  \label{fig:flowers21_o}
  }
  \hskip -2.2mm
  \subfigure[Saliency]{
  \includegraphics[width=0.21\linewidth]{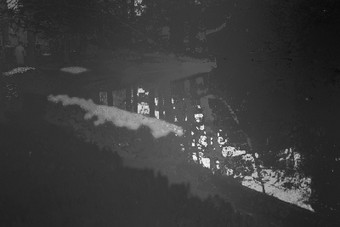}
  \label{fig:flowers21_sal}
  }
  \hskip -2.2mm
  \subfigure[Ours]{
  \includegraphics[width=0.105\linewidth]{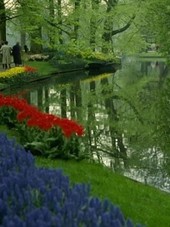}
  \label{fig:flowers21_tex}
  }
  \hskip -2.2mm
  \subfigure[HSD Saliency]{
  \includegraphics[width=0.21\linewidth]{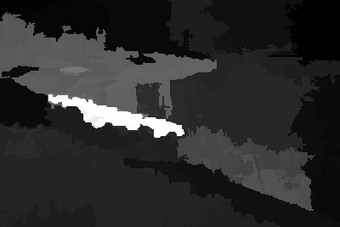}
  \label{fig:flowers21_sal_hsd}
  }
  \hskip -2.2mm
  \subfigure[Ours by HSD]{
  \includegraphics[width=0.105\linewidth]{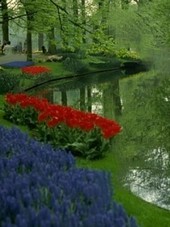}
  \label{fig:fflowers21_tex_hsd}
  }
  \hskip -2.2mm
  \subfigure[BDS]{
  \includegraphics[width=0.105\linewidth]{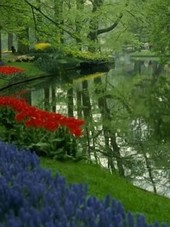}
  \label{fig:flowers21_bd}
  }
  \vskip -3mm
 \caption{\protect \small Comparison of retargeting results by using our saliency map and by using the one of HSD method~\protect\cite{Yan:2013:HSD}. More semantic informations (e.g. people and yellow flowers) are kept in our result while better preserving the boundary continuity of the grassland.}
  \label{fig:flowers21}
  \vskip -3mm
\end{figure*}

\vspace{-4mm}
\subsection{Texture-Based Significance Map Generation}
\label{sec:texsal}

\vspace{-2mm}
Pixel significance measurements have been commonly used in image retargeting approaches. Usually saliency map is employed to help generate the significance map of the input image. However, the purpose of almost all of current saliency detection algorithms is to detect and segment the distinct salient objects. \textit{To the best of our knowledge, a saliency detection algorithm aiming at marking the visually important areas (not objects) of a texture has not been proposed.} We present a hierarchical framework to deal with saliency detection of a texture. We first segment the image into $M$ patches by using the SLIC method~\cite{Achanta:2012:SSC}. Since a texture usually does not contain a distinct salient object, the saliency detection becomes determining the patches which are visually unique from others. Previous approaches usually use color or contrast information to evaluate the visual difference between pixels or patches~\cite{Cheng:2011:GCB,Margolin:2013:WMP}, but this is not effective enough for dealing with texture images, as shown in Fig.~\ref{fig:field01}. In our approach, in order to better evaluate the saliency of a texture, we integrate 2D Gabor filter~\cite{Pang:2013:FGT} with 4 frequencies and 6 directions to extract the texture features of the T-regions. For each SLIC patch $A_i$, we calculate the average and variance of Gabor values of all the pixels in it and then get a 48D texture feature. Thus, we define the visual uniqueness saliency cue of $A_i$ as a weighted sum of color difference and texture difference from other patches:
%
\begin{eqnarray}
U_i = \sum_{j = 1}^{M}({w(A_i) \cdot \exp(\frac{-D_s(A_i, A_j)}{\sigma_s^2})} \\ \nonumber
\cdot (\|\textbf{C}_i - \mathbf{C}_j\|^2 + \|\textbf{G}_i - \textbf{G}_j\|^2)),
\end{eqnarray}
where $\textbf{C}_i$/$\textbf{C}_j$ is the average color of a patch, $\textbf{G}_i$/$\mathbf{G}_j$ is the texture feature. The color feature and texture feature are both normalized to $[0,1]$. $w(A_i)$ counts the number of T-pixels in $A_i$. Patches with more T-pixels contribute higher visual uniqueness weights than those containing less T-pixels. $D_s(A_i, A_j)$ is the square of Euclidean distance between patch centroids of $A_i$ and $A_j$, and $\sigma_s$ controls the strength of spatial weighting. In our implementation, we set $\sigma_s^2 = 0.5$ with pixel coordinates normalized to $[0, 1]$.

Similar as \cite{Yan:2013:HSD}, we also add the location heuristic that in many cases pixels close to a natural image center could be salient:
$$
H_i = \frac{1}{w(A_i)}\sum_{x_j \in A_i}{\exp(-\lambda\|\mathbf{x}_j - \mathbf{x}_c\|^2)},
$$
where $\mathbf{x}_j$ is the coordinate of a pixel in patch $A_i$, and $\mathbf{x}_c$ is the coordinate of image center. In our experiments, we set $\lambda = 9$ to balance the visual uniqueness and location cues. We combine $H_i$ with $U_i$ to get the saliency of patch $A_i$:
$$
S_i = U_i \cdot H_i.
$$
For further robustness, we compute patch-based saliency at three scales: $M = 100, 500, 1000$ and average them pixel by pixel. As shown in Fig.~\ref{fig:clover01_sal_texcol}, we can get a coarse saliency map by using the above patch-based hierarchical method. Finally, we adopt an image up-sampling method~\cite{Criminisi:2010:GIV} to refine the coarse saliency map and assign a saliency value to each image pixel. We define the saliency $\tilde{S}_i$ of a pixel as a Gaussian weighted linear combination of the saliency of its $N$ neighbourhoods:
$$
\tilde{S}_i = \frac{1}{Z_i}\sum_{j = 1}^{N}{\exp(-\frac{\|\mathbf{c}_i - \mathbf{c}_j\|^2 + \|\mathbf{g}_i - \mathbf{g}_j\|^2 + \|\mathbf{x}_i - \mathbf{x}_j\|^2}{2\sigma})S_j},
$$
where $\mathbf{c}_i$ is the pixel color, $\mathbf{g}_i$ is the Gabor texture feature, and $\mathbf{x}_i$ is the pixel coordinate. We set $\sigma = 30$ in all our experiments. Result is shown in Fig.~\ref{fig:clover01_sal}. Similar refinement method is used in \cite{Perazzi:2012:SFC}, but they only consider the color and position features.
\begin{figure*}[t!]
\centering
\subfigure[Original]{
  \includegraphics[width=0.26\linewidth]{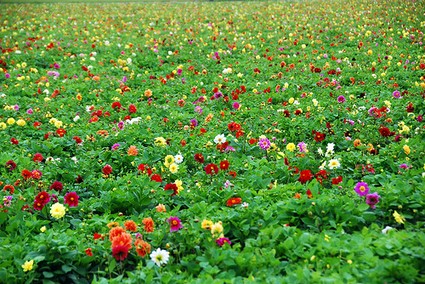}
  \label{fig:flowers13_o}
  }
  \hskip -2mm
  \subfigure[Ours]{
  \includegraphics[width=0.1352\linewidth]{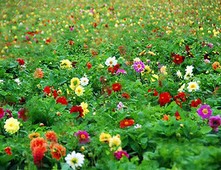}
  \label{fig:flowers13_texd}
  }
  \hskip -2mm
  \subfigure[Inverse~\protect\cite{Wei:2008}]{
  \includegraphics[width=0.1352\linewidth]{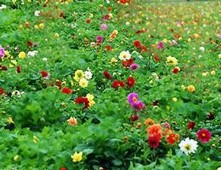}
  \label{fig:flowers13_its}
  }
  \hskip -2mm
  \subfigure[Optimization~\protect\cite{Kwatra:2005}]{
  \includegraphics[width=0.1352\linewidth]{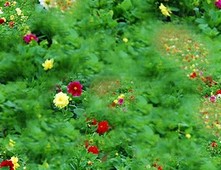}
  \label{fig:flowers13_to}
  }
  \hskip -2mm
  \subfigure[Appearance~\protect\cite{Lefebvre:2006:ATS}]{
  \includegraphics[width=0.1352\linewidth]{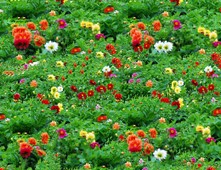}
  \label{fig:flowers13_ats}
  }
  \hskip -2mm
  \subfigure[BDS~\protect\cite{Simakov:08}]{
  \includegraphics[width=0.1352\linewidth]{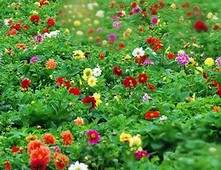}
  \label{fig:flowers13_bds}
  }
  \vskip -3mm
 \caption{\protect \small Comparison of previous normal texture synthesis methods to our method. Our method can both preserve the perspective effect and the salient areas. Input resolution $500\times 334$, output resolution $260 \times 200$. $49.09\%$ users favour our result.}
 \vskip -3mm
  \label{fig:flowers13}
\end{figure*}

As discussed above, previous saliency detection approaches are usually designed to highlight the salient object(s). As shown in Fig.~\ref{fig:clover01}, for an image in our dataset, previous methods either over-darken or over-highlight most part of a T-region. They also have difficulties with accurately detecting the visually important areas of T-regions due to the lack of texture features. On the other hand, the content balance will be easily damaged during retargeting if the saliency values of T-regions are too smaller or too larger than NT-regions, especially when the sizes of T- and NT-regions are similar. Therefore, to address these problems, in the saliency map of the whole image, we replace the parts of T-regions with the saliency maps generated by our method. For the generation of initial saliency map, we use the method in \cite{Yan:2013:HSD} if the area of NT-region is less than $30\%$ of the image since this method is good at distinguishing salient objects from complex background patterns. Otherwise, we use HSD method~\cite{Yang:2013:SDG} to generate a more balanced initial saliency map. We use the saliency map as the significance map for retargeting operation. In Fig.~\ref{fig:flowers21} we show an example of using different saliency maps to retarget an image. We can see that our method can highlight more visually unique contents in the saliency map than HSD.

\vspace{-4mm}
\subsection{Initial Retargeting}
\label{sec:initret}

\vspace{-2mm}
As an initial retargeting operation, we first smooth the original image by structure extraction~\cite{Xu:2012:SET}. We then use F-MultiOp method~\cite{Dong:2012} to resize the smoothed image to the target size. The significance map is utilized to preserve the important areas of both T- and NT-regions. The operation details are recorded, including the numbers of the three operators (i.e., seam carving, homogeneous scaling, and cropping) and the paths of pixels used by seam carving. Finally the original image is retargeted by copying these operations. This scheme can efficiently eliminate the unexpected affects of large-magnitude gradients of complex texture details to seam carving. After initial retargeting, the resized NT-region will be directly used in the final result, but we re-generate the T-regions by content-aware synthesis.

\vspace{-4mm}
\subsection{Content-Aware Synthesis for T-regions}
\label{sec:syn}

\vspace{-2mm}
We synthesize a T-region of the resized image by using the original T-region as the example. However, for most images, directly synthesizing the content by normal texture synthesis algorithms cannot generate satisfied result or even change the semantics of the image and introduce obvious boundary discontinuity. The global visual appearance may be damaged when the resized ratio is large. As shown in Figs.~\ref{fig:flowers13_its}-\ref{fig:flowers13_ats}, the perspective characteristic no longer exists and the spatial structure of the content is also damaged. For our content-aware image resizing application, the synthesis algorithm should preserve the globally visual appearances of the original T-regions as well as the local continuity.

\textbf{Initialization} We employ patch-based synthesis framework which is effective for image textures to synthesize the resized T-regions. In our experiments, we find that a \emph{good} initialization will increase the quality of the resized results. Therefore, we use the resized T-regions generated by F-MultiOp in initial retargeting as the initial guess. With the help of significance map during F-MultiOp, this will effectively preserve the global visual appearance and the visually salient areas in the result.
\begin{figure}[h]
\centering
\subfigure[Original Image]{
  \includegraphics[width=0.46\linewidth]{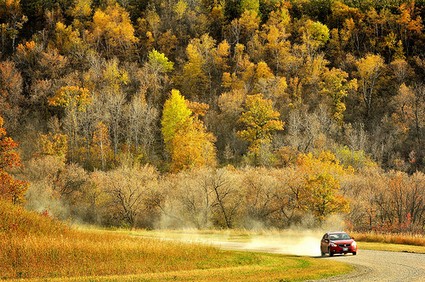}
  \label{fig:car_o}
  }
  \hskip -2mm
  \subfigure[Ours with $\mu_p$]{
  \includegraphics[width=0.23\linewidth]{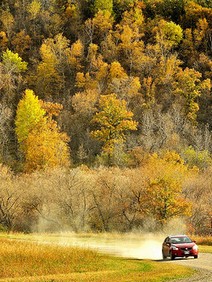}
  \label{fig:car_texd}
  }
  \hskip -2mm
  \subfigure[\scriptsize Ours without $\mu_p$]{
  \includegraphics[width=0.23\linewidth]{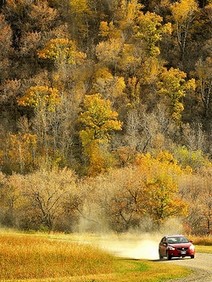}
  \label{fig:car_texd_rep}
  }
  \vskip -3mm
  \caption{\protect \small The artifact of repeat patterns can be avoided by increasing the neighborhood matching cost with a penalty coefficient.}
  \label{fig:car}
  \vskip -3mm
\end{figure}
\begin{figure*}
\centering
\subfigure[Original Image]{
  \includegraphics[width=0.3\linewidth]{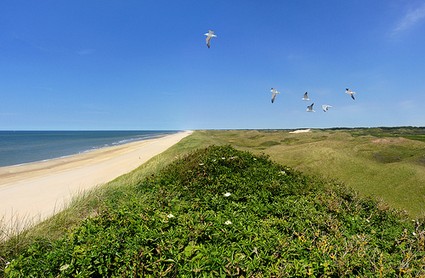}
  \label{fig:field18_o}
  }
  \hskip -1mm
  \subfigure[Discontinuity]{
  \includegraphics[width=0.156\linewidth]{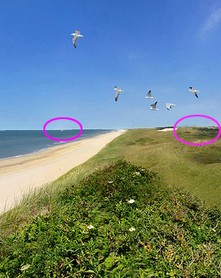}
  \label{fig:field18_texd_nf}
  }
  \hskip -1mm
  \subfigure[Fix discontinuity]{
  \includegraphics[width=0.156\linewidth]{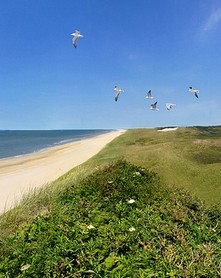}
  \label{fig:field18_texd}
  }
  \hskip -1mm
  \subfigure[AAD]{
  \includegraphics[width=0.156\linewidth]{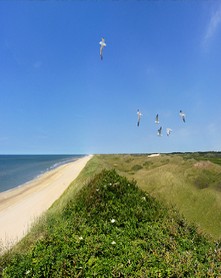}
  \label{fig:field18_aad}
  }
  \hskip -1mm
  \subfigure[BDS]{
  \includegraphics[width=0.156\linewidth]{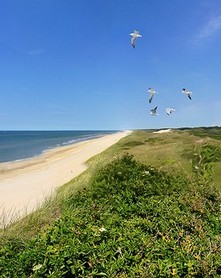}
  \label{fig:field18_bd}
  }
  \vskip -3mm
  \caption{\protect \small Reducing the discontinuity artifact by growing the T-region boundaries in the original image, so as to provide overlapping regions for determining the seamless cut-path in the result. Input $500\times 327$, output $260 \times 333$. $52.73\%$ users favour our result.}
  \label{fig:field18}
  \vskip -4mm
\end{figure*}

\textbf{Neighborhood metric} The neighborhood similarity metric is the core component of example-based texture synthesis algorithms~\cite{Wei:2009}. We denote $Z_p$ as the spatial neighborhood around a sample $p$, which is constructed by taking the union of all pixels within its spatial extent defined by a user specified neighborhood size. We formulate the distance metric between the neighborhoods of two sample $p$ and $q$ as:
\begin{equation}
M(Z_p; Z_q) = \mu_p \cdot (\sum_{p' \in Z_p}{\|\mathbf{c}_{p'} - \mathbf{c}_{q'}\|^2} + \omega \cdot \|\mathbf{x}_{p'} - \mathbf{x}_{q'}\|^2),
\label{equ:nmetric}
\end{equation}
where $p'$  runs through all pixels $\in Z_p$, $q' \in Z_q$ is the spatially corresponding sample of $p'$, $\mathbf{c}$ represents the pixel color in RGB space, and $\mathbf{x}$ is the local coordinate of a sample pixel. Different from traditional texture synthesis that usually defines the neighborhood metric as a simple sum-of-squared of the pixel attributes (such as colors and edges), we add the spatial information to the neighborhood metric. The spatial item can preserve the global appearance without causing over-smoothing and generating obvious partial/broken objects (detailedly discussed in Sect.~\ref{sec:results}). In Equation~(\ref{equ:nmetric}), $\mu_p$ is a penalty coefficient which is used to avoid overusing the same patches in the resulting image:
$$
\mu_p = 1 + \beta \cdot t_p,
$$
where $t_p$ is number of times that patch $Z_p$ has been used in the resulting image, $\beta = 10$ is a constant. In Fig.~\ref{fig:car}, we can see the importance of adding $\mu_q$ to the neighborhood metric in avoiding unexpected repeat patterns. Note that for Fig.~\ref{fig:car_texd_rep} we also did not integrate the significance map during the initial retargeting process, so the salient yellow trees in the middle of the original image are lost in the result.

\textbf{Optimization} Therefore, given an original exemplar T-region $\mathcal{I}$, our goal is to synthesize an output $\mathcal{O}$ that contains similar visual appearances to $\mathcal{I}$. We formulate this as an optimization problem via the following energy function:
\begin{equation}
E(\mathcal{I}; \mathcal{O}) = \sum_{p \in \mathcal{O}}{M(Z_p; Z_q)},
\label{equ:energy}
\end{equation}
where the first term measures the similarity between the input exemplar $\mathcal{I}$ and $\mathcal{O}$ via our local neighborhoods metric as defined in Equation~(\ref{equ:nmetric}). Specifically, for each output sample $q \in \mathcal{O}$, we find the corresponding input sample $p \in \mathcal{I}$ with the most similar neighborhood (according to Equation~(\ref{equ:nmetric})), and sum their squared neighborhood differences. Our goal is to find an output $\mathcal{O}$ with a low energy value. For our normal image resizing applications, we assume as null. Furthermore, we follow the EM-like methodology in \cite{Kwatra:2005} to optimize Equation~(\ref{equ:energy}) because of its high quality and generality with different boundary conditions.

We perform our synthesis process in multi-resolutions through an iterative optimization solver. For Equation~(\ref{equ:nmetric}), we use larger $\omega$ in lower resolution to increase the spatial constraint. This scheme helps to preserve the global appearance during the synthesis process, then we decrease the $\omega$ value in higher resolution to avoid the local texel repeat. In all our experiment, we use a 3-level pyramid and within each level, from lower to higher, we fix $\omega = 0.65, 0.25, 0.1$.

\textbf{Adaptive neighborhood matching} In each iteration, we search for the most similar input neighborhood for each output sample and assign the exemplar patch from the matched neighborhood to the output. This will gradually improve the synthesis quality. During the search step, exhaustively examining every input sample to minimize the energy value in Equation~(\ref{equ:nmetric}) can be computationally expensive. Previous works use $K$-means~\cite{Kwatra:2005} or $K$-coherence~\cite{Han:2006} to find an approximate nearest neighborhood (ANN). These strategies can efficiently accelerate the search process. However, when the texel diversity increases, the ANNs may not be accurate enough to improve the neighborhood quality, which will cause dissatisfied results (Fig.~\ref{fig:field01_texd_k}). Therefore, we also search for the exact nearest neighborhoods by brute-force method over the exemplar image. Since the nearest neighborhoods are independent from each other, we implement our EM-based synthesis algorithm fully on GPU by implementing the search in a parallel framework, which will dramatically accelerate the search process. Specifically, in each thread, we calculate the similarity of two neighborhoods in the M-step and perform the average operation for each pixel in the E-step. Moreover, to further accelerate the neighborhood matching process, we use an adaptive scheme to narrow the searching domain in finer layers. Since our synthesis algorithm is a hierarchical framework which contains three layers, we use an adaptive scheme to gradually narrow the searching domain. In layer 1 where the images are processed in the lowest resolution, we search the best patch from the whole exemplar for each patch in the resulting image, we search for the best matching from the whole exemplar. Then, in layer 2, for each patch in the resulting image, we narrow the searching domain to the $40\%$ pixels of the exemplar around its corresponding patch. Furthermore, we narrow the searching domain to $20\%$ in the finest layer. Note that in the two finer layers, we still perform full search in the first matching operation and narrow the domain in the latter steps.

\textbf{Synthesis as a whole} We synthesize the image as a whole when most contents of the scene are textures (usually more than $70\%$). The advantage of this strategy is that it can better preserves the global visual appearance and effectively reduce the object broken artifacts. Fig.~\ref{fig:tire}, \ref{fig:car}, \ref{fig:flowers22}, and \ref{fig:village} show 4 typical examples of this class. Our results are directly generated by the synthesis operator.

\vspace{-4mm}
\subsection{Merge of T-Regions and NT-Regions}
\label{sec:merge}

\vspace{-2mm}
Since we resize the T- and NT- regions by different strategy, there may exist discontinuity of image contents between them. As demonstrated in Fig.~\ref{fig:field18}, the image content on the boundary between T- and NT-regions may be changed after synthesis. To reduce the discontinuity artifact, we grow the boundary by expanding 4 pixels on both inward and outward sides. We then get an overlapping area between the T- and NT-regions in the resized image. Afterwards, we re-synthesize those boundaries pixels by using the original image as the input example. The inclusion of NT- pixels on the boundary in the synthesis process helps to maintain the content consistency. Fig.~\ref{fig:field18_texd_nf} and \ref{fig:field18_texd} compare the results without and with fixing the discontinuity, respectively.

\vspace{-4mm}
\section{Results and Discussion}
\label{sec:results}

\vspace{-2mm}
We have implemented our method on a PC with Intel Core(TM) i7 950 CPU, 3.06 GHz, 8GB RAM, and nVidia Geforce GTX 770 GPU with 2048MB video memory. Our T-region synthesis algorithm is fully implemented on GPU with CUDA. The texture detection and saliency detection are both performed in real-time. The timing of resizing examples shown in this paper ranges from 10 seconds to 40 seconds, depending on the sizes of the output T-regions.

Figs.~\ref{fig:tulip04}, \ref{fig:tire}, \ref{fig:field01}-\ref{fig:flowers13}, and \ref{fig:field18}-\ref{fig:village} show our image retargeting results. We perform a user study for visual comparison (detailedly described below). For each figure, we put our result and primarily better than other results with relatively higher votes. We can see that our content-aware synthesis method can preserve the overall texture features in terms of texel shape, perspective, boundary continuity, content completeness, and clarity. The perspective appearance remains perspective. The shapes of texels are reasonably preserved, without over-squeezing/over-stretching or uneven distortion of texels within the regions. The boundary between T- and NT-region is continuous. All the prominent contents of textures appear in the result.

\vspace{-4mm}
\subsection{Evaluation on Textural Scene Retargeting Dataset}

\begin{figure}[h]
\centering
\subfigure[Original]{
  \includegraphics[width=0.3\linewidth]{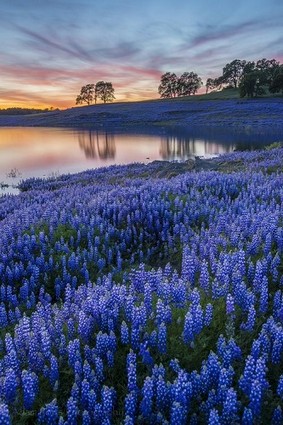}
  \label{fig:lavender_o}
  }
  \hskip -1.5mm
  \begin{minipage}[b]{0.65\linewidth}
  \subfigure[Ours]{
  \includegraphics[width=0.462\linewidth]{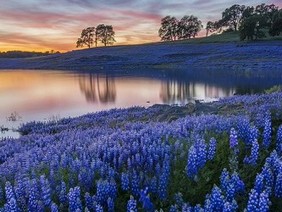}
  \label{fig:lavender_texd}
  }
  \hskip -1.5mm
  \subfigure[AAD]{
  \includegraphics[width=0.462\linewidth]{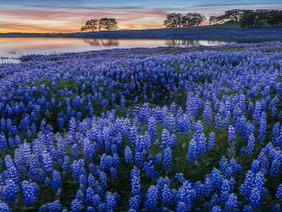}
  \label{fig:lavender_aad}
  }

  \vskip -1mm
  \subfigure[BDS]{
  \includegraphics[width=0.462\linewidth]{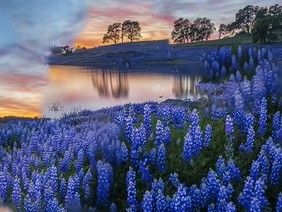}
  \label{fig:lavender_bd}
  }
  \hskip -1.5mm
  \subfigure[Cropping]{
  \includegraphics[width=0.462\linewidth]{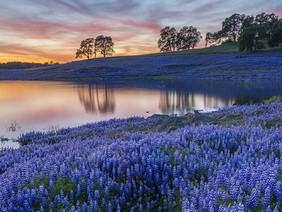}
  \label{fig:lavender_cr}
  }
  \end{minipage}
  \vskip -1mm
  
  \subfigure[F-MultiOp]{
  \includegraphics[width=0.3\linewidth]{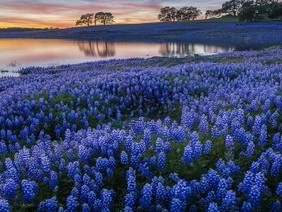}
  \label{fig:lavender_mod}
  }
  \hskip -1.5mm
  \subfigure[PBW]{
  \includegraphics[width=0.3\linewidth]{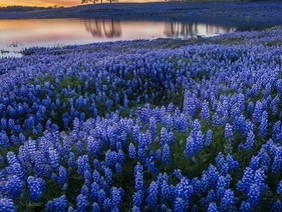}
  \label{fig:lavender_pbw}
  }
  \hskip -1.5mm
  \subfigure[Shift-Map]{
  \includegraphics[width=0.3\linewidth]{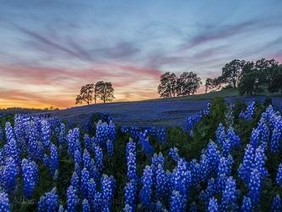}
  \label{fig:lavender_sm}
  }
  \vskip -3mm
  
 \caption{\protect \small Visual comparison. Our method better preserves the scene layout and the perspective effect of the lavender. Perspective is lost in the cropping result. $67.27\%$ users favour our result.}
 \vskip -6mm
  \label{fig:lavender}
\end{figure}

\vspace{-2mm}
Although images from RetargetMe benchmark~\cite{Rubinstein:2010} have a large variety in their content, most of their textural regions are simple and smooth. To represent more general situations that real world images fall into, we construct a Textural Scene Retargeting Dataset (TSRD) with 61 images. They all contain diversified and large textural patterns (occupying more than $50\%$ areas of the whole image). These images are collected from RetargetMe (9 images), CSSD~\cite{Yan:2013:HSD} (Fig.~\ref{fig:flowers21_o}) and internet. Some images in the RetargetMe benchmark which also contain textures are not included in TSRD because either the T-regions are small or the textures are relatively smooth without obvious texels (such as a still water surface, a smooth snowfield, and a manicured lawn). The images in the new dataset can be roughly divided into three types: pure textures with vivid global visual effects (Type 1), images with textures around one or more salient objects (Type 2), images with distinct T- and NT-regions (Type 3). For the exemplars in this paper, Figs.~\ref{fig:clover01_o}, \ref{fig:flowers21_o}, \ref{fig:flowers13_o}, and \ref{fig:flowers22_o} belong to Type 1. Figs.~\ref{fig:tire_o}, \ref{fig:car_o}, \ref{fig:child_o}, and \ref{fig:girl02_o} belong to Type 2. Figs.~\ref{fig:tulip04}(a), \ref{fig:field13_tex_o}, \ref{fig:field01_o}, \ref{fig:car_o}, \ref{fig:field18_o}, \ref{fig:lavender_o}, \ref{fig:field19_o}, \ref{fig:field14_o}, \ref{fig:bicycle1_o} and \ref{fig:village_o} belong to Type 3. The whole dataset and the comparisons with previous state-of-the-arts methods are all shown in the supplemental material.

\vspace{-4mm}
\subsection{Comparison with previous methods}

\vspace{-2mm}
For quantitative evaluation, we compare our method with six state-of-the-arts image retargeting approaches, i.e., Axis-Aligned Deformation (AAD)~\cite{Panozzo:2012}, Bi-Directional Similarity (BDS)~\cite{Simakov:08}, Cropping, Multi-Operator (F-MultiOp~\cite{Dong:2012} and MultiOp~\cite{Rubinstein:09}), Patch-Based Warping (PBW)~\cite{Lin:2013:PBI} and Shift-Map~\cite{Pritch:09}. The experiments are performed on our data set.
\begin{figure*}[t!]
\centering
\subfigure[Original Image]{
  \includegraphics[width=0.268\linewidth]{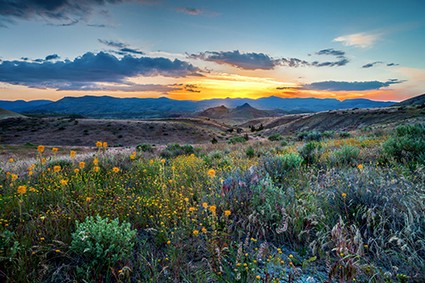}
  \label{fig:field19_o}
  }
  \hskip -2mm
  \subfigure[Ours]{
  \includegraphics[width=0.134\linewidth]{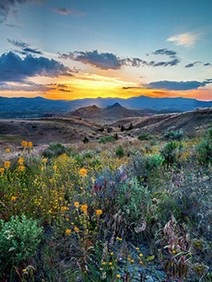}
  \label{fig:field19_texd}
  }
  \hskip -2mm
  \subfigure[BDS]{
  \includegraphics[width=0.134\linewidth]{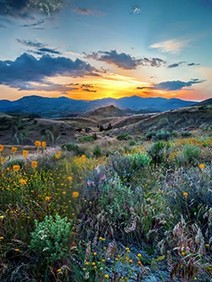}
  \label{fig:field19_bd}
  }
  \hskip -2mm
  \subfigure[F-MultiOp]{
  \includegraphics[width=0.134\linewidth]{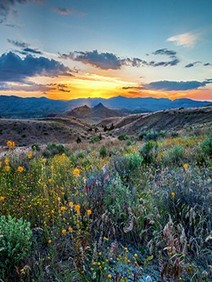}
  \label{fig:field19_mod}
  }
  \hskip -2mm
  \subfigure[PBW]{
  \includegraphics[width=0.134\linewidth]{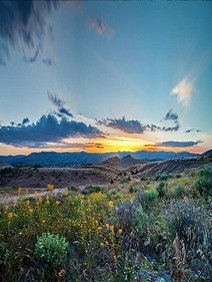}
  \label{fig:field19_pbw}
  }
  \hskip -2mm
  \subfigure[Shift-Map]{
  \includegraphics[width=0.134\linewidth]{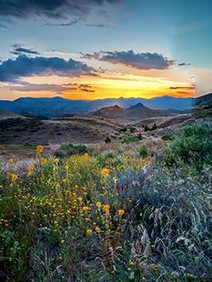}
  \label{fig:field19_sm}
  }
 \vskip -3mm
 \caption{\protect \small Input resolution $500 \times 333$. Target resolution $250 \times 333$. $60.00\%$ users favour our result.}
  \label{fig:field19}
  \vskip -4mm
\end{figure*}
\begin{figure*}[t!]
\centering
\subfigure[Original Image]{
  \includegraphics[width=0.26\linewidth]{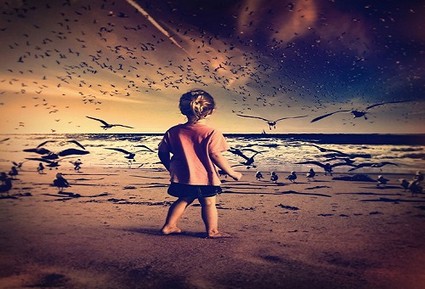}
  \label{fig:child_o}
  }
  \hskip -2mm
  \subfigure[Ours]{
  \includegraphics[width=0.1352\linewidth]{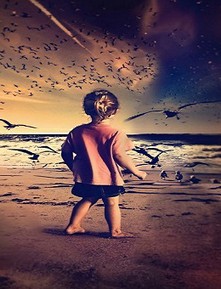}
  \label{fig:child_texd}
  }
  \hskip -2mm
  \subfigure[BDS]{
  \includegraphics[width=0.1352\linewidth]{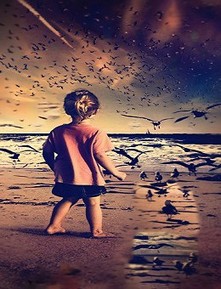}
  \label{fig:child_bds}
  }
  \hskip -2mm
  \subfigure[F-MultiOp]{
  \includegraphics[width=0.1352\linewidth]{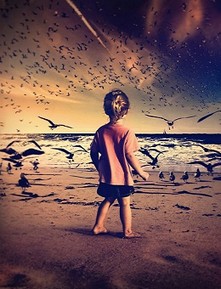}
  \label{fig:child_mod}
  }
  \hskip -2mm
  \subfigure[PBW]{
  \includegraphics[width=0.1352\linewidth]{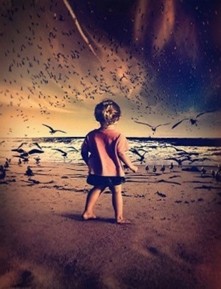}
  \label{fig:child_pbw}
  }
  \hskip -2mm
  \subfigure[Shift-Map]{
  \includegraphics[width=0.1352\linewidth]{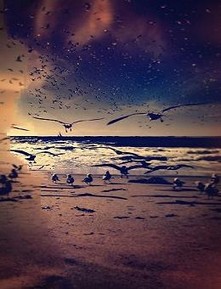}
  \label{fig:child_sm}
  }
  \vskip -3mm
 \caption{\protect \small Input resolution $500 \times 340$. Target resolution $260 \times 340$. $43.64\%$ users favour our result.}
  \label{fig:child}
  \vskip -4mm
\end{figure*}
\begin{figure*}[t!]
\centering
\subfigure[Original Image]{
  \includegraphics[width=0.268\linewidth]{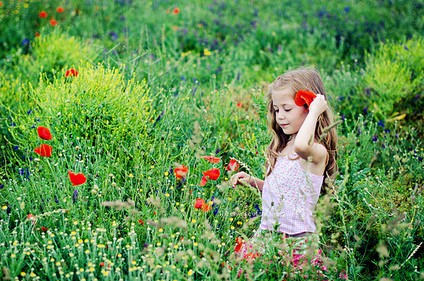}
  \label{fig:girl02_o}
  }
  \hskip -2mm
  \subfigure[Ours]{
  \includegraphics[width=0.134\linewidth]{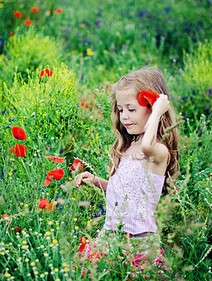}
  \label{fig:girl02_texd}
  }
  \hskip -2mm
  \subfigure[BDS]{
  \includegraphics[width=0.134\linewidth]{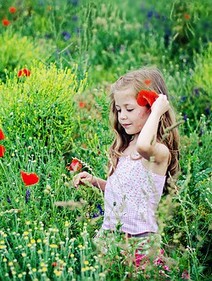}
  \label{fig:girl02_bd}
  }
  \hskip -2mm
  \subfigure[F-MultiOp]{
  \includegraphics[width=0.134\linewidth]{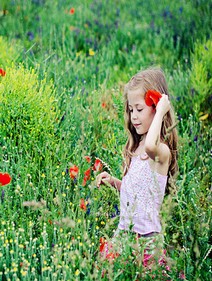}
  \label{fig:girl02_mod}
  }
  \hskip -2mm
  \subfigure[PBW]{
  \includegraphics[width=0.134\linewidth]{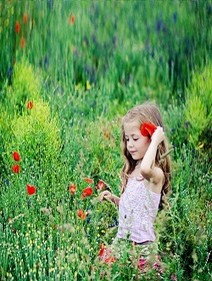}
  \label{fig:girl02_pbw}
  }
  \hskip -2mm
  \subfigure[Shift-Map]{
  \includegraphics[width=0.134\linewidth]{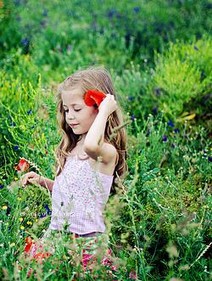}
  \label{fig:girl02_sm}
  }
  \vskip -3mm
 \caption{\protect \small Input resolution $500 \times 331$. Target resolution $250 \times 331$. $61.82\%$ users favour our result.}
  \label{fig:girl02}
  \vskip -4mm
\end{figure*}
\begin{figure*}[t!]
\centering
\subfigure[Original Image]{
  \includegraphics[width=0.268\linewidth]{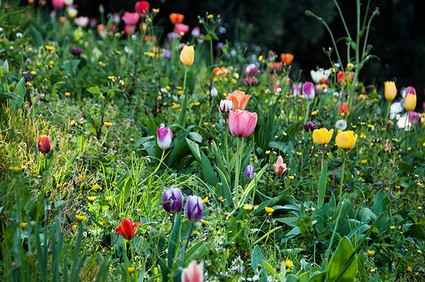}
  \label{fig:flowers22_o}
  }
  \hskip -2mm
  \subfigure[Ours]{
  \includegraphics[width=0.134\linewidth]{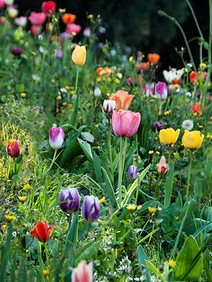}
  \label{fig:flowers22_texd}
  }
  \hskip -2mm
  \subfigure[AAD]{
  \includegraphics[width=0.134\linewidth]{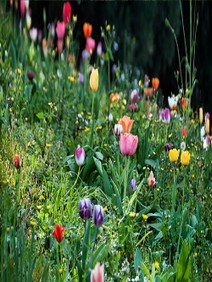}
  \label{fig:flowers22_aad}
  }  
  \hskip -2mm
  \subfigure[BDS]{
  \includegraphics[width=0.134\linewidth]{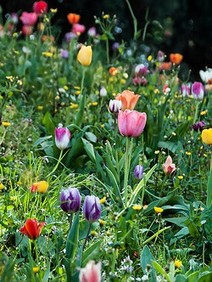}
  \label{fig:flowers22_bd}
  }
  \hskip -2mm
  \subfigure[F-MultiOp]{
  \includegraphics[width=0.134\linewidth]{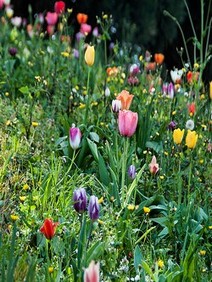}
  \label{fig:flowers22_mod}
  }
  \hskip -2mm
  \subfigure[Shift-Map]{
  \includegraphics[width=0.134\linewidth]{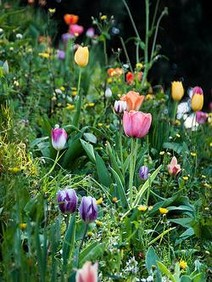}
  \label{fig:flowers22_sm}
  }
  \vskip -3mm
 \caption{\protect \small Input resolution $500 \times 332$. Target resolution $250 \times 332$. $49.09\%$ users favour our result.}
  \label{fig:flowers22}
  \vskip -4mm
\end{figure*}
\begin{figure*}[t!]
\centering
\subfigure[Original Image]{
  \includegraphics[width=0.26\linewidth]{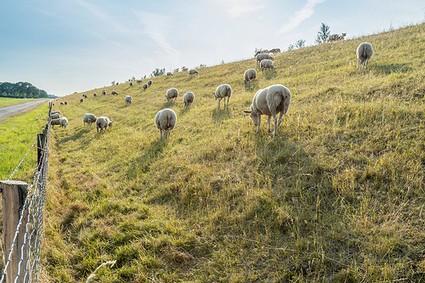}
  \label{fig:field14_o}
  }
  \hskip -2mm
  \subfigure[Ours]{
  \includegraphics[width=0.1352\linewidth]{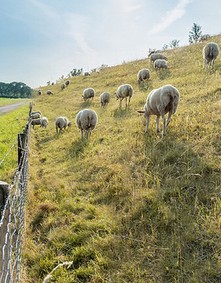}
  \label{fig:field14_texd}
  }
  \hskip -2mm
  \subfigure[AAD]{
  \includegraphics[width=0.1352\linewidth]{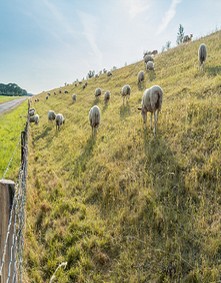}
  \label{fig:field14_aad}
  }
  \hskip -2mm
  \subfigure[BDS]{
  \includegraphics[width=0.1352\linewidth]{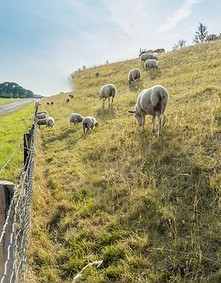}
  \label{fig:field14_bd}
  }
  \hskip -2mm
  \subfigure[Cropping]{
  \includegraphics[width=0.1352\linewidth]{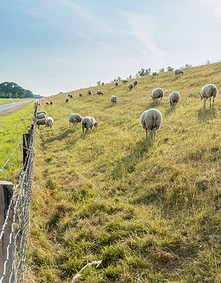}
  \label{fig:field14_cr}
  }
  \hskip -2mm
  \subfigure[Shift-Map]{
  \includegraphics[width=0.1352\linewidth]{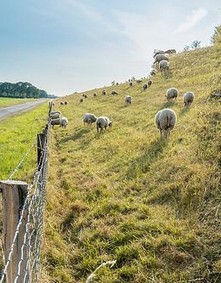}
  \label{fig:field14_sm}
  }
 \vskip -3mm
 \caption{\protect \small Input resolution $500 \times 333$. Target resolution $260 \times 333$. $52.73\%$ users favour our result.}
  \label{fig:field14}
  \vskip -4mm
\end{figure*}
\begin{figure*}[t!]
\centering
\subfigure[Original]{
  \includegraphics[width=0.268\linewidth]{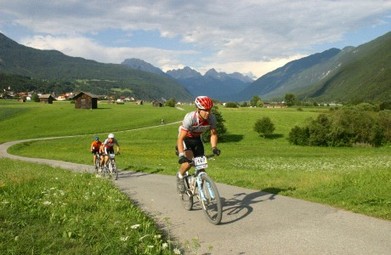}
  \label{fig:bicycle1_o}
  }
  \hskip -2mm
  \subfigure[Ours]{
  \includegraphics[width=0.134\linewidth]{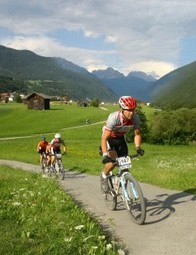}
  \label{fig:bicycle1_texd}
  }
  \hskip -2mm
  \subfigure[Cropping]{
  \includegraphics[width=0.134\linewidth]{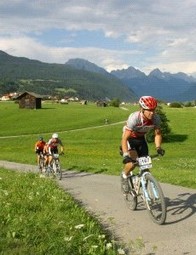}
  \label{fig:bicycle1_cr}
  }  
  \hskip -2mm
  \subfigure[MultiOp]{
  \includegraphics[width=0.134\linewidth]{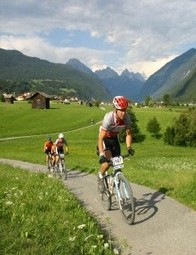}
  \label{fig:bicycle1_mod}
  }
  \hskip -2mm
  \subfigure[PBW]{
  \includegraphics[width=0.134\linewidth]{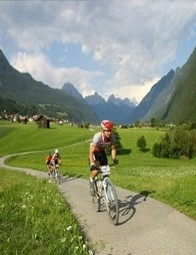}
  \label{fig:bicycle1_pbw}
  }
  \hskip -2mm
  \subfigure[Shift-Map]{
  \includegraphics[width=0.134\linewidth]{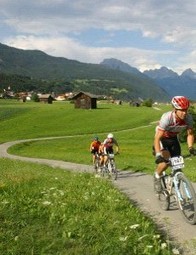}
  \label{fig:bicycle1_sm}
  }
  \vskip -3mm
 \caption{\protect \small Input resolution $460 \times 300$. Target resolution $230 \times 300$. $43.64\%$ users favour our result.}
  \label{fig:bicycle1}
  \vskip -4mm
\end{figure*}
\begin{figure*}[t!]
\centering
\subfigure[Original Image]{
  \includegraphics[width=0.26\linewidth]{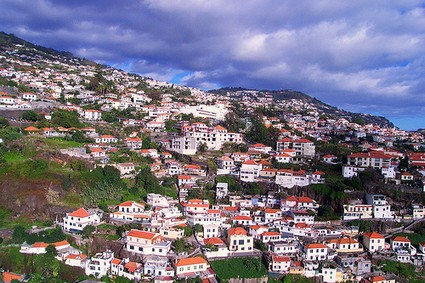}
  \label{fig:village_o}
  }
  \hskip -2mm
  \subfigure[Ours]{
  \includegraphics[width=0.1352\linewidth]{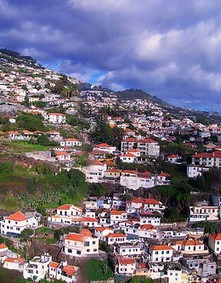}
  \label{fig:village_texd}
  }
  \hskip -2mm
  \subfigure[AAD]{
  \includegraphics[width=0.1352\linewidth]{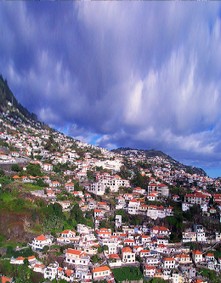}
  \label{fig:village_aad}
  }
  \hskip -2mm
  \subfigure[BDS]{
  \includegraphics[width=0.1352\linewidth]{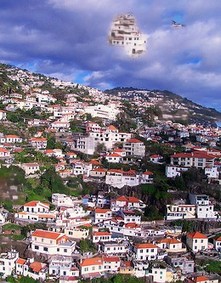}
  \label{fig:village_bds}
  }
  \hskip -2mm
  \subfigure[F-MultiOp]{
  \includegraphics[width=0.1352\linewidth]{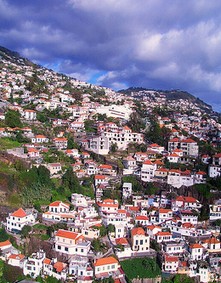}
  \label{fig:village_mod}
  }
  \hskip -2mm
  \subfigure[Shift-Map]{
  \includegraphics[width=0.1352\linewidth]{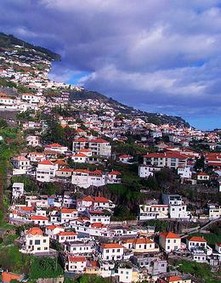}
  \label{fig:village_sm}
  }
  \vskip -3mm
 \caption{\protect \small Input resolution $500 \times 333$. Target resolution $260 \times 333$. $52.73\%$ users favour our result.}
  \label{fig:village}
  \vskip -4mm
\end{figure*}

\textit{\textbf{For AAD and PBW}}, we choose them for comparison since they are two typical continuous image warping approaches, which have been recently presented and testified to be among the best warping methods. SV~\cite{Krahenbuhl:2009} is also a good warping method which has been proved by the test on RetargetMe benchmark. However, in \cite{Panozzo:2012} the user study demonstrates that AAD is better than SV, so we only compare our method with AAD and PBW. The AAD results are generated with authors' program by using the default parameters. The PBW results are provided by the original author. When dealing with images in TSRD, compared to our method, the main problem of AAD and PBW is in many cases they will over-squeeze some contents (e.g., Figs.~\ref{fig:tulip04}(e), \ref{fig:field23_pbw}, \ref{fig:lavender_aad} and \ref{fig:village_aad}) or the salient objects (e.g., Figs.~\ref{fig:tire_aad}, \ref{fig:child_pbw} and \ref{fig:bicycle1_pbw}), while over-stretch the background (Figs.~\ref{fig:field19_pbw} and \ref{fig:girl02_pbw}), which makes some visually important regions to be too small in the resulting images. In many results the content structures of the scenes are obvious imbalance. The main reason is because warping usually tends to maintain as many as contents while preserving the aspect ratios of the areas with large energy or significance values. In most images of TSRD, these areas are usually the T-regions (Type 3) or the salient objects (Type 2). Therefore, to maintain the shape of those "important" areas, we can find that in the results generated by AAD or PBW, the T-regions are either overstretched (e.g., Fig.~\ref{fig:girl02_pbw}) or over-squeezed (e.g., Fig.~\ref{fig:village_aad}). Uneven distortion to the salient objects may also appear if their significance values are low, such as Figs.~\ref{fig:tire_aad} and \ref{fig:field14_aad}. Specifically, as shown in Fig.~\ref{fig:flowers22_aad}, when the scene is almost all constructed by textures (Type 1), all the contents maybe be distorted if we use warping-based methods.

\begin{figure}
\includegraphics[width=0.98\linewidth]{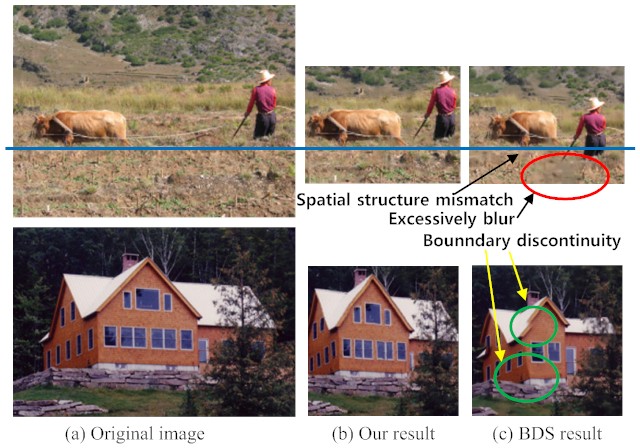}
\vskip -2mm
\caption{\protect \small The BDS results often suffer artifacts of excessively blur, spatial structure mismatch and boundary discontinuity.}
\vskip -4mm
\label{fig:bdstexdcomp}
\end{figure}

\textit{\textbf{For BDS}}, we choose it for comparison since it is a synthesis-based image summarization method. The results are generated by imagestack program (\url{http://code.google.com/p/imagestack/}). For each exemplar, we use different parameters to generate four images and manually choose the best one as the final result. At each gradually resizing step, we set the EM iteration times as 50 and refinement interation times for each intermediate target as 20. When dealing with the images in TSRD, compared to our method, the main problem of BDS is that there will be obvious boundary discontinuity, such as the mountain in Fig.~\ref{fig:tulip04}(c), the beach in Fig.~\ref{fig:field18_bd}, the sky in Fig.~\ref{fig:lavender_bd}, and the grassland in Fig.~\ref{fig:field14_bd}. The reason is that BDS only uses color distance for neighbourhood matching, while the integration of spatial information in our algorithm can ensure the content continuity. The second problem often appears in BDS is the over-smoothing of some areas, such as the left-bottom tulips in Fig.~\ref{fig:tulip04}(c), the middle of the bough and the bottom of the trunk in Fig.~\ref{fig:tire_bd}, and the small yellow flowers in Fig.~\ref{fig:field19_bd}. We consider that it is due to the strategy of bidirectional similarity, sometimes one area in the resulting image is "obliged" to be similar as multiple areas of the original image. Our single-directional framework can avoid this problem. In fact, for image retatgeting application, content loss is allowed. Most of the users will be satisfied if the important contents are preserved. Another common problem of BDS is spatial structure mismatch of content. That is, some patches may appear in wrong places, such as the mountain patches in the sky of Fig.~\ref{fig:tulip04}(c), the flowers patches in the sky of Fig.~\ref{fig:field23_bd}, and the house patches in the sky of Fig.~\ref{fig:village_bds}. The other phenomenon of this problem is spatial relationship of some contents may be wrong in the result, such as the child and the seabirds in Fig.~\ref{fig:child_bds}, and the farm cattle and the farmer in Fig.~\ref{fig:bdstexdcomp}(c) (the the farmer should be above the blue line). We consider that this is also due to the lack of spatial constraint in the synthesis algorithm. We use Fig.~\ref{fig:bdstexdcomp} to show the main problems of BDS, the two exemplars are picked from the original paper~\cite{Simakov:08}. Moreover, missing a good significance map also causes the loss of visually important contents in the results, such as the missing of salient red flowers in Fig.~\ref{fig:girl02_bd} and yellow flowers in Fig.~\ref{fig:flowers22_bd}. Our good saliency map also makes it enough for our optimization process to only use a single-directional neighbourhood matching since the important areas are preserved in the initial retargeting operation. This also efficiently accelerates the speed of the synthesis process. In our experiments, we find that BDS usually costs more than 20 minutes to generate a good result, which limits its practical use in many applications. On the other hand, \textit{\textbf{PatchMatch}} method~\cite{Barnes:09} can also perform image retargeting by synthesis. We do not compare with PatchMatch since it shares the same framework as BDS so that it can be treated as a parallel method for image retargeting.

\textit{\textbf{For Cropping}}, we choose it for comparison since in most cases it is the first choice of the users during the comparative study of \cite{Rubinstein:2010}. On the other hand, a texture usually appears a certain self-similarity, so maybe a simple cropping will be enough to well summarize its content. The results are created by an expert photographer. When dealing with images in TSRD, compared to our method, the main problem of cropping is some important contents will be unavoidably lost if there are multiple important contents located near the different sides of the input image, such as the largest sheep in Fig.~\ref{fig:field14_cr}, and the trees and mountain in Fig.~\ref{fig:bicycle1_cr}. Our synthesis strategy can narrow the distance between the important contents and make them to appear together in the result. Moreover, as discussed in \cite{Rubinstein:2010}, cropping should be considered as a reference, not as a proper retargeting algorithm. Here we still decide to compare with cropping only because sometimes it can benefit from the self-similarity characteristic of some textures and generate good retargeting results.
\begin{figure*}[t!]
\centering
\subfigure[Original Image]{
  \includegraphics[width=0.268\linewidth]{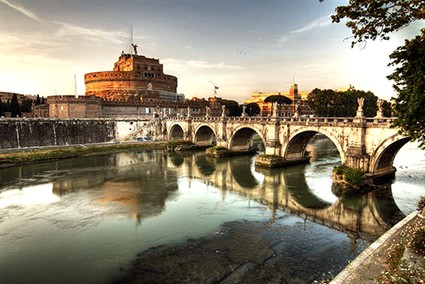}
  \label{fig:setangle_o}
  }
  \hskip -2mm
  \subfigure[Ours]{
  \includegraphics[width=0.134\linewidth]{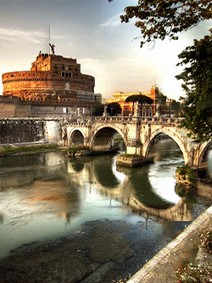}
  \label{fig:setangle_texd}
  }
  \hskip -2mm
  \subfigure[AAD]{
  \includegraphics[width=0.134\linewidth]{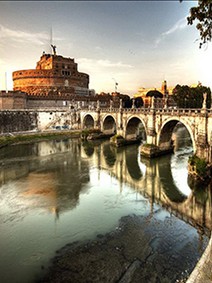}
  \label{fig:setangle_aad}
  }  
  \hskip -2mm
  \subfigure[BDS]{
  \includegraphics[width=0.134\linewidth]{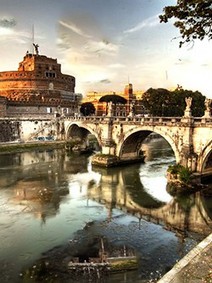}
  \label{fig:setangle_bd}
  }
  \hskip -2mm
  \subfigure[MultiOp]{
  \includegraphics[width=0.134\linewidth]{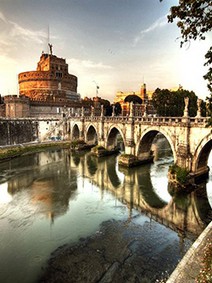}
  \label{fig:setangle_mod}
  }
  \hskip -2mm
  \subfigure[Shift-Map]{
  \includegraphics[width=0.134\linewidth]{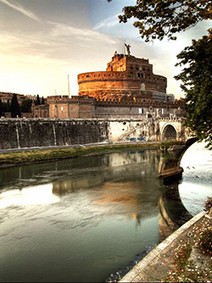}
  \label{fig:setangle_sm}
  }
  \vskip -3mm
 \caption{\protect \small Retargeting a general image with our method. Our method achieves a summarization-like result. The bridge, is shortened.}
  \label{fig:setangle}
  \vskip -4mm
\end{figure*}
\begin{figure*}[t!]
\centering
\subfigure[Original Image]{
  \includegraphics[width=0.2\linewidth]{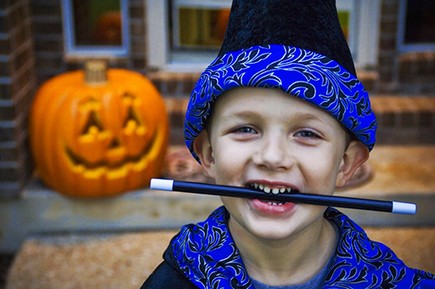}
  \label{fig:child02_o}
  }
  \hskip -2.5mm
  \subfigure[Ours]{
  \includegraphics[width=0.15\linewidth]{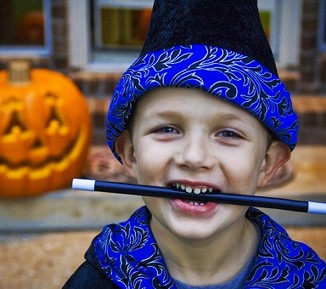}
  \label{fig:child02_texd}
  }
  \hskip -2.5mm
  \subfigure[AAD]{
  \includegraphics[width=0.15\linewidth]{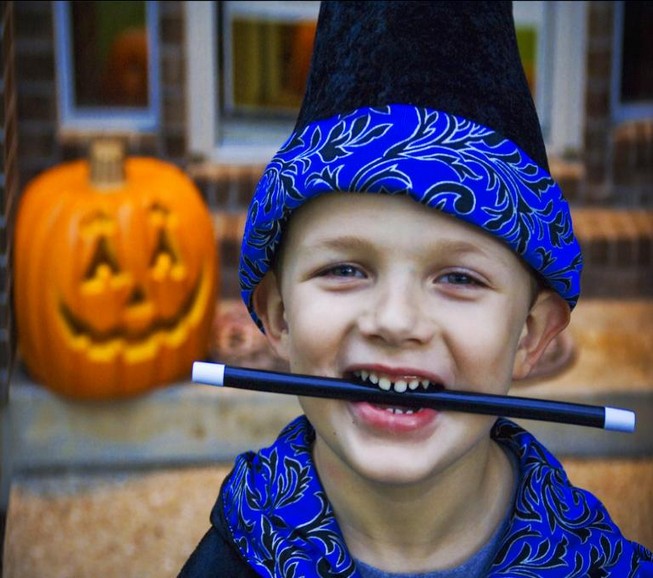}
  \label{fig:child02_aad}
  }  
  \hskip -2.5mm
  \subfigure[BDS]{
  \includegraphics[width=0.15\linewidth]{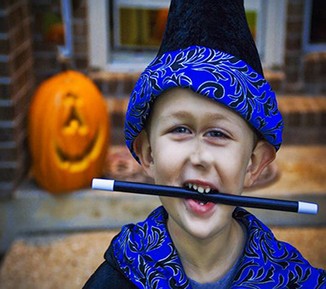}
  \label{fig:child02_bd}
  }
  \hskip -2.5mm
  \subfigure[MultiOp]{
  \includegraphics[width=0.15\linewidth]{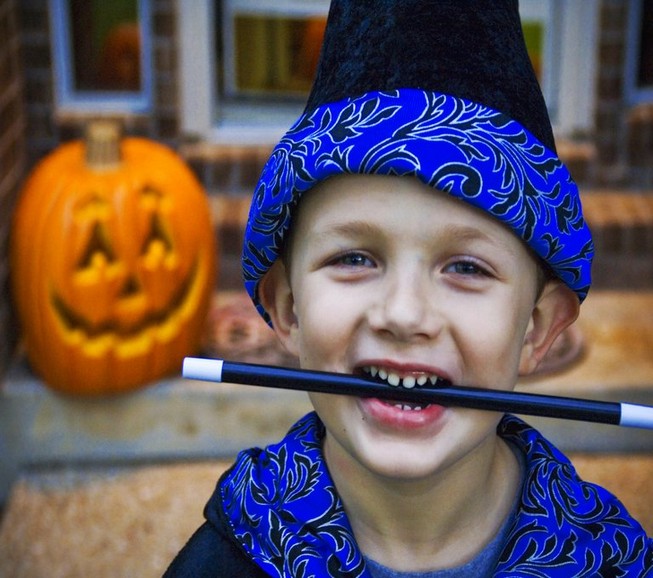}
  \label{fig:child02_mod}
  }
  \hskip -2.5mm
  \subfigure[Shift-Map]{
  \includegraphics[width=0.15\linewidth]{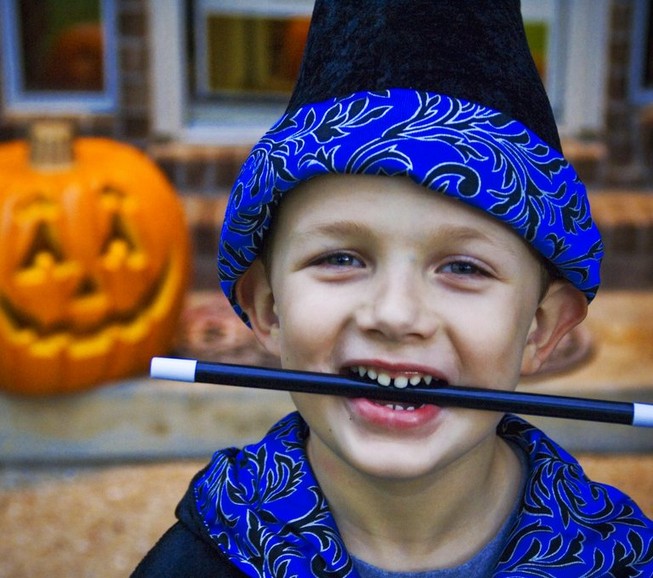}
  \label{fig:child02_sm}
  }
  \vskip -3mm
 \caption{\protect \small Retargeting a general image with our method. Our synthesis operator can also handle some non-textural images.}
  \label{fig:child02}
  \vskip -5mm
\end{figure*}

\textit{\textbf{For F-MultiOp and MultiOp}}, we choose them for comparison since the MultiOp framework outperforms most algorithms according to the comparative study~\cite{Rubinstein:2010}. F-MultiOp method has been demonstrated in \cite{Dong:2012} that it can generate results of the similar quality as MultiOp, so we consider these two methods as the same in our comparison. The MultiOp results of the six images collected from RetargetMe benchmark are directly downloaded from the AAD website (\url{http://igl.ethz.ch/projects/retargeting/aa-retargeting/aa-comparisons/dataset/index.html}), including the AAD results of those six images. The other results are generated by using F-MultiOp, which are all provided by the original author. When dealing with images in TSRD, compared to our method, the main problem of multi-operator methods is the uneven distortion to objects or texels, such are the tulips in Fig.~\ref{fig:tulip04}(d), the tire in Fig.~\ref{fig:tire_mod}, the girl and flowers in \ref{fig:girl02_mod}, the flowers in \ref{fig:flowers22_mod}, and the sportsman in Fig.~\ref{fig:bicycle1_mod}. The main reason is because although the integration of cropping operator can somewhat avoid the overall distortion, the unavoidable use of seam carving and homogeneous scaling operators (to protect the similarity between original image and resulting image) may still cause uneven distortions to objects or texels, especially when the T-regions are distributed throughout one dimension of the original image, such as Figs.~\ref{fig:tulip04}(a), \ref{fig:field19_o}, and \ref{fig:flowers22_o}. This problem can only be solved by using a synthesis-based strategy. On the other hand, as shown in Fig.~\ref{fig:lavender_mod}, some important contents may be over-squeezed due to the lack of a good significance map.

\vspace{-2mm}
\textit{\textbf{For Shift-Map}}, we choose it for comparison since sometimes it can generate a synthesis-like result which selectively stitches some contents together to construct a resized image. The results are partly provided by the original author, partly generated with the authors' online system, and partly generated with a public implementation (\url{https://sites.google.com/site/shimisalant/home/image-retargeting}) after the online system is taken down. When dealing with images in TSRD, compared to our method, the main problem of Shift-Map is that in many cases it will unpredictably lose some important contents (e.g., the left cherry tree in Fig.~\ref{fig:field23_sm}, the river in Fig.~\ref{fig:lavender_sm}, the child in Fig.~\ref{fig:child_sm}, and the largest sheep in Fig.~\ref{fig:field14_sm}) or degenerate to cropping which will damage the composition of the original image (e.g., the girl's location is too left in Fig.~\ref{fig:girl02_sm}, and the sportsman's location is too right in Fig.~\ref{fig:bicycle1_sm}). The main reason is because stitching are minimized due to the global smoothness term. On the other hand, to get a good retargeting result by using Shift-Map, sometimes the user need to gradually resize the image by manually setting the number of removed columns/rows. This strategy is useful in preserving salient objects in the resulting image but ineffective for our images because for textures it usually does not contain distinct long boundaries that can help to penalize the removal of a large area. We consider that this is just the reason that in some cases shift-map degenerates to cropping when dealing with our images. Another problem of shift-map method is that it may also cause boundary discontinuity artifacts, such as the string of the tire in Fig.~\ref{fig:tire_sm}, the grassland boundary in Fig.~\ref{fig:field14_sm}, and the mountain boundary in Fig.~\ref{fig:village_sm}.

\vspace{-1mm}
\textit{\textbf{For example-based texture synthesis}}, apparently the normal texture synthesis algorithms such as texture optimization~\cite{Kwatra:2005} and appearance-space texture synthesis~\cite{Lefebvre:2006:ATS} are not fit for image retargeting since they are originally designed for enlargement but have no effective schemes for size decrease. Inverse texture synthesis (ITS)~\cite{Wei:2008} can produce a small texture compaction that  summarizes the original. Its framework is very similar as BDS method so it will suffer the same problems as BDS if being used for image retargeting. On the other hand, the textural contents in most our images are not standard textures so using pure texture synthesis framework will easily cause content discontinuity or damage the globally varying effects.

\vspace{-1mm}
In Figs.~\ref{fig:setangle} and \ref{fig:child02}, we show two examples of only using our synthesis operator to retarget a general image which does not contain obvious textural patterns. Results show that our synthesis operator can also works well for some general images. However, since our synthesis operator is specifically designed for dealing with textural patterns, we can not assure of synthesizing satisfied results for arbitrary non-textural images. In fact in our framework, the NT-region is retargeted by F-MultiOp instead of the synthesis operator.

\vspace{-4mm}
\subsection{User Study}

\begin{figure}[h]
\includegraphics[width=0.98\linewidth]{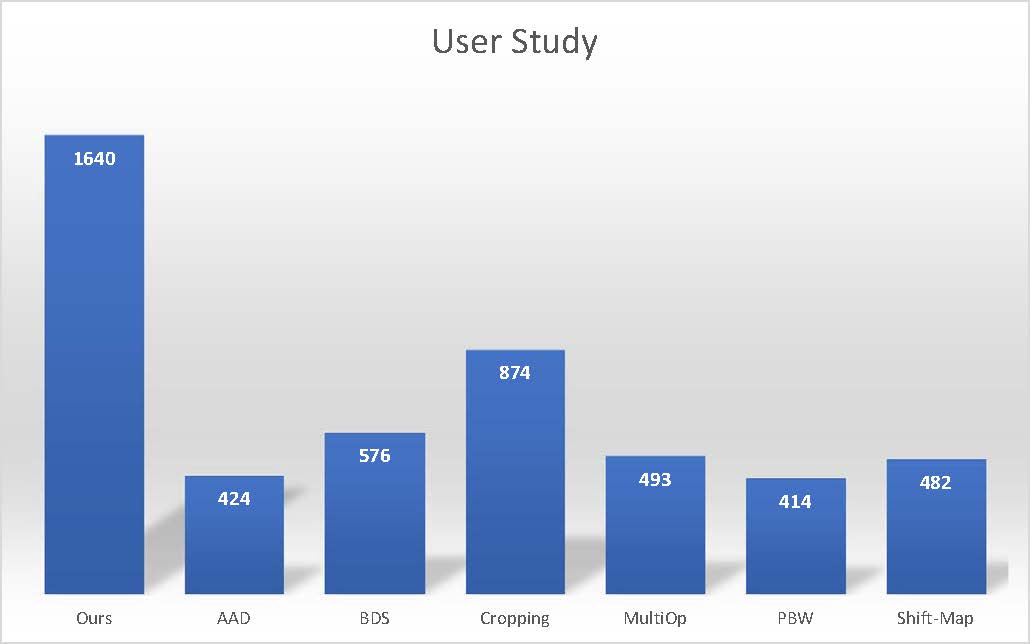}
\caption{\protect \small The statistics of the user study result.}
\vspace{-4mm}
\label{fig:user}
\end{figure}

\vspace{-2mm}
To evaluate our method further, we perform a user study to compare the results from different methods. All the stimuli are shown in the supplemental material. A total of 55 participants (24 males, 21 females, age range 20-45) from different backgrounds attended the comparison of 61 sets of resized images. Each participant is paid \$10 for their participation. All the participants sat in front of a 22-inch computers of $1680 \times 1050$ px in a semi-dark room. In the experiment, we showed the original image, our result, and the images of the competitors. We then ask which image the participant prefers. For each group, the original image is separately shown in the first row, while the results are randomly displayed in two additional rows within the same page. We allow the participant to choose at most two favourite images from the results. We did not provide a time constraint for the decision time. However, we recommend for the participants to finish the tests within 30 min. We allow the participants to move back and forth across the different pages by clicking the mouse. The average finishing time is 26 min 53 sec. A total of 4903 votes are reported. Fig.~\ref{fig:user} shows the the statistics of how many times the results of each method has been chosen as favourite retargeting results. Based on the statistics, our method outperforms all competitors in general. For each test exemplar in TSRD, we show the percentages when our method and the competitors have been chosen by the participants in the supplemental material.

\vspace{-4mm}
\subsection{Limitations}

\begin{figure}[h]
\centering
 \vspace{-3mm}
\subfigure[Original Image]{
  \includegraphics[width=0.4\linewidth]{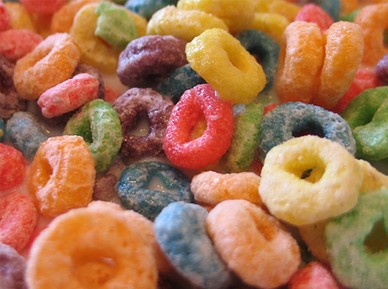}
  \label{fig:candy_o}
  }
  \subfigure[Ours]{
  \includegraphics[width=0.2281\linewidth]{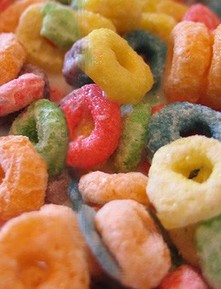}
  \label{fig:candy_texd}
  }
  \subfigure[Shift-Map]{
  \includegraphics[width=0.2281\linewidth]{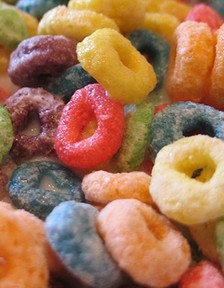}
  \label{fig:candy_sm}
  }
  \vskip -3mm
 \caption{\protect \small Input resolution $456 \times 340$. Target resolution $264 \times 340$. $29.09\%$ users favour our result.}
 \vspace{-3mm}
  \label{fig:candy}
\end{figure}

\vspace{-2mm}
The main limitation of our algorithm is the speed. Although we implement our synthesis operator fully on GPU, we still cannot get real-time performance like most warping-based methods, especially when the T-regions are large. Our method may generate unsatisfied results when the texels are very large (like an object) and have different attributes (color, shape, orientation, etc.). Fig.~\ref{fig:candy} shows such one example, the texels (a candy) are large and visually different from each other. Therefore, we can see that there are obvious object discontinuity in our result. In this case, one possible way to improve retargeting quality is to use object carving~\cite{Dong:2014:SBI} to entirely remove some objects.

\vspace{-4mm}
\section{Conclusion and Future Work}
\label{sec:con}

\vspace{-2mm}
The scenes containing textural regions are very common in natural images. However, as shown in our paper, most of them cannot be well handled by current general image resizing algorithms due to the lack of high level semantic information. We introduces a novel concept and robust method to solve the problem. An automatic methodology is proposed to detect the textures and adjust the saliency information. Then we use a synthesis-based image resizing system to achieve natural resizing effects with minimum texel visual appearance damage. The integration of the spatial information ensures the content consistency between the original image and the result images. Our spatial-aware strategy can be integrated into most existing general resizing frameworks and enhance their robustness. Experiments shown that our system can handle a great variety of input scenes especially non-standard textural regions (for example Fig.~\ref{fig:car_o} is combined with many separate textural objects). For future work, extending the example-based synthesis operator to 3D scene resizing can be an interesting direction.



\ifCLASSOPTIONcaptionsoff
  \newpage
\fi

\bibliographystyle{IEEEtran}
\bibliography{Photo}








\end{document}